\documentclass[preprintnumbers, floatfix, showkeys, preprintnumbers, letterpaper, twocolumn, superscriptaddress,nofootinbib]{revtex4-2}
\pdfoutput=1 
\usepackage{yfonts}
\usepackage{bm}
\usepackage{graphicx}
\usepackage{microtype}
\usepackage{amsmath}
\usepackage{amsthm}
\usepackage{amssymb}
\usepackage{subcaption}
\usepackage{url}
\usepackage{hyperref}
\usepackage{mathtools}
\usepackage{orcidlink}
\usepackage{slashbox}
\usepackage{booktabs}
\usepackage{relsize}
\usepackage{color}
\usepackage{mathrsfs}
\usepackage{calrsfs}
\usepackage{amsfonts}
\usepackage{tabularx}
\usepackage{latexsym}
\usepackage{ragged2e}
\usepackage{epsfig}
\usepackage{textcomp}
\usepackage{float}

\newtheorem*{theorem*}{Theorem}

\newtheorem*{proposition*}{Proposition}
\makeatletter
\def\@opargbegintheorem#1#2#3{\trivlist
   \item[]{\bfseries #1\ #2\ (#3)} \itshape}
\makeatother

\usepackage{textcomp}
\usepackage[most]{tcolorbox}
\usepackage{tikz}
\usepackage[object=vectorian]{pgfornament}
\usetikzlibrary{calc}

\usepackage{caption}
\DeclareCaptionJustification{justified}{\leftskip=0pt \rightskip=0pt \parfillskip=0pt plus 1fil}
\definecolor{vividviolet}{rgb}{0.62, 0.0, 1.0}
\definecolor{amaranth}{rgb}{0.9, 0.17, 0.31}
\definecolor{palatinateblue}{rgb}{0.15, 0.23, 0.89}
\definecolor{brightpink}{rgb}{1.0, 0.0, 0.5}
\definecolor{cornflowerblue}{rgb}{0.39, 0.58, 0.93}
\definecolor{deepcarminepink}{rgb}{0.94, 0.19, 0.22}
\definecolor{radicalred}{rgb}{1.0, 0.21, 0.37}

\hypersetup{ linktoc=all,
	colorlinks, linkcolor={palatinateblue},
	citecolor={brightpink}, urlcolor={amaranth}
}

\graphicspath{{Images/}}

\renewcommand{\d}[1]{\ensuremath{\operatorname{d}\!{#1}}}

\newcommand{\SBH}{S_{\rm BH}}
\newcommand{\mpl}{M_{\rm Pl}}

\newcommand{\Ord}{\mathcal{O}}

\def\sideremark#1{\ifvmode\leavevmode\fi\vadjust{\vbox to0pt{\vss
			\hbox to 0pt{\hskip\hsize\hskip1em
				\vbox{\hsize1.3cm\tiny\raggedright\pretolerance10000
					\noindent #1\hfill}\hss}\vbox to8pt{\vfil}\vss}}}%
\def\beq{\begin{equation}}
\def\eeq{\end{equation}}

\newtcolorbox{manuscripteq}{
  enhanced,
  colback=white,
  colframe=black,
  boxrule=0.8pt,
  arc=0pt,
  left=8pt,
  right=8pt,
  top=8pt,
  bottom=8pt,
  borderline={0.4pt}{3pt}{black},
}

\newcommand{\vintageeq}[1]{%
\begin{tikzpicture}[baseline=(EQ.base)]

\node[
    inner xsep=25pt,
    inner ysep=16pt
] (EQ) {$\displaystyle #1$};

\draw[line width=.55pt]
  ($(EQ.north west)+(-.22,.20)$)
  rectangle
  ($(EQ.south east)+(.22,-.20)$);

\node[anchor=north west]
  at ($(EQ.north west)+(-.22,.20)$)
  {\pgfornament[width=.72cm]{63}};

\node[anchor=north east]
  at ($(EQ.north east)+(.22,.20)$)
  {\pgfornament[width=.72cm,symmetry=v]{63}};

\node[anchor=south west]
  at ($(EQ.south west)+(-.22,-.20)$)
  {\pgfornament[width=.72cm,symmetry=h]{63}};

\node[anchor=south east]
  at ($(EQ.south east)+(.22,-.20)$)
  {\pgfornament[width=.72cm,symmetry=c]{63}};

\end{tikzpicture}%
}

\begin{document}
\title{The Forgetful Black Hole:\\
Memory Burden and Hysteresis in Hawking Evaporation}

\author{Yen Chin \surname{Ong}\orcidlink{0000-0002-3944-1693}}
\email{ongyenchin@nuaa.edu.cn}
\affiliation{Center for the Cross-disciplinary Research of Space Science and Quantum-technologies (CROSS-Q), College of Physics, Nanjing University of Aeronautics and Astronautics, \\29 Jiangjun Road, Nanjing City, Jiangsu Province 211106, China}

\begin{abstract}
The memory-burden scenario of black hole evaporation proposes that Hawking emission may become increasingly suppressed as maintaining stored quantum information becomes dynamically costly. In this work, we revisit this scenario and investigate the conditions under which information retention can produce a genuine dynamical burden. We derive general scaling conditions for incremental memory accumulation at a fixed fractional mass loss and establish a conditional no-go result: entropy-suppressed, zero-mean deposits with subextensive integrated correlations cannot produce an order-one luminosity modification through unamplified accumulation and bounded feedback. We identify three essential requirements for dynamical memory: a record of the past must be written, retained, and coupled back to subsequent evaporation. We illustrate these requirements with concrete toy models that exhibit either a prolonged but finite black hole lifetime or an asymptotically approached finite-mass state (an effective remnant). We further distinguish kinetic-memory, thermodynamic-memory, and spectral-memory sub-categories of possible memory-burden mechanisms. In particular, we demonstrate that an independently evolving memory variable can generate evaporation--accretion hysteresis, such that a black hole returning to the same mass may possess a different retained memory and consequently a different luminosity. This history dependence may have implications for primordial black hole phenomenology. Finally, we briefly discuss the possibility of ``merger rejuvenation'', in which black hole mergers reorganize or reset the effective memory state.
\end{abstract}

\maketitle
\section{Introduction: Black Hole Evaporation and Memory-Burden}

In the standard picture\footnote{We shall work in the natural units $G=\hbar=c=k_B=1$.}, the Bekenstein-Hawking entropy and the Hawking temperature for asymptotically flat Schwarzschild black holes are, respectively,
\begin{equation}
  \SBH=\frac{A}{4}=4\pi M^2,\qquad
  T_H=\frac{1}{8\pi M},
\end{equation}
where $A$ is the horizon area and $M$ the mass of the black hole.
The associated Hawking luminosity is given by the standard Stefan-Boltzmann law $L_H \propto AT_H^4$, which leads to the scaling
\begin{equation}
  L_H\equiv-\dot M=\frac{\alpha}{M^2},
  \label{eq:HawkingFlux}
\end{equation}
where $\alpha$ encodes particle content and greybody factors. The evaporation time scale, in the simplest model that assumes geometrical optic approximation, is obtained by integrating Eq.\eqref{eq:HawkingFlux}, which gives
\begin{equation}
  t_\text{life}\sim M_0^3.
  \label{eq:tev}
\end{equation}
Note that two Schwarzschild black holes with the same instantaneous mass possess the same temperature and the same Hawking luminosity, independently of how they reached that mass. 

However, the ``memory burden'' scenario  \cite{Dvali2018,DvaliEtAl2020,DvaliEtAl2024} could substantially modify this picture. Essentially, the idea is that black holes possess highly occupied ``master modes'' that support a large set of memory modes. The master modes are like the macroscopic order parameter, while the memory modes are the constituent degrees of freedom. The master modes interact with the
memory modes in an attractive manner. As the master modes deplete during the course of Hawking evaporation, maintaining the stored memory pattern becomes increasingly costly, and this causes the evaporation to slow down, and may even eventually halt altogether. While the final state of the black hole would depend on the exact details of the model, the general message is that if indeed there exists memory burden, then a drastic departure from the standard Hawking evaporation should be expected no later than when the black hole has lost an order-one fraction of its mass. The stronger claim is that this should happen at the latest after losing half of the mass \cite{DvaliEtAl2020}. A subsequent work explored some similarities between the evaporation of black holes and saturated solitons in quantum field theory. The findings suggested that the information retrieval process can be strongly influenced by the memory burden effect \cite{ContriDvaliSakhelashvili2025}. 

The idea of memory burden has received much attention especially in the theoretical astrophysics community, particularly in primordial black hole (PBH) cosmology. For example, memory-burdened evaporation has been used to reopen light PBH dark matter windows \cite{ThossBurkertKohri2024, AlexandreDvaliKoutsangelas2024,2503.21005,2506.13861,2503.21740,2512.06941,2604.06858} (see, however, \cite{2503.21005}); to study gravitational waves \cite{KohriTeradaYanagida2024,2409.04436,2510.19916}, high-scale leptogenesis \cite{CalabreseChianeseSaviano2025}, 21 cm cosmology \cite{2512.06941}, Big Bang nucleosynthesis (BBN) constraints \cite{ChaudhuriKohriThoss2025,2606.04707}, high energy gamma rays \cite{Chianese2025,TanZhou2026} and cosmic rays \cite{2603.15827}, accretion-assisted survival and dark radiation production \cite{Maity2026}, non-cold dark matter \cite{2604.00090}, and thermal hot spots generated by evaporating PBHs \cite{LevyHeurtier2025}.
Meanwhile, IceCube and other neutrino data \cite{2410.07604,2606.16145,2608.08144}, the DAMPE data of electron measurements \cite{2503.13192}, and graviton-photon conversion rate \cite{2511.01848} were used to constrain memory-burdened PBHs. 
Extension to regular black hole physics was recently considered in \cite{DuZhangLi2026,2608.21353}.

While the astrophysical works explore the phenomenological implications of memory-burdened Hawking evaporation, the post-semiclassical suppression is often assumed without further theoretical understanding.
In fact, comparatively little attention has been devoted to model-independent dynamical criteria for when stored information actually modifies the evaporation law. 
This issue has become especially sharp with the recent counterexample of Banks \cite{Banks2026}.
Within the Holographic Space-Time (HST) models of black hole evaporation, Banks argued that a large information capacity and long-lived constrained degrees of freedom need not produce memory burden\footnote{The argument combines Fermi's Golden Rule and detailed balance with HST-specific notions of energy, causal diamonds, and the rapidly increasing phase space available to fragmented outgoing states. We do not need the details in this work.}.
It should be emphasized that, although HST is a specifically holographic construction and as such need not be directly applicable to astrophysical black holes, it nevertheless provides an explicit example in which microscopic information can remain encoded while the black hole still does not experience memory burden.
Thus the current theoretical landscape now contains: (1) explicit constructions intended to produce a memory burden, and (2) an explicit construction in which information is stored but nevertheless does not burden the decay.

Given the ongoing strong interest in memory burden in the astrophysics community, it is therefore timely for us to ask the obvious question: \emph{What are the necessary ingredients for any mechanism that produces an order-one history-dependent modification of Hawking evaporation at a fixed, order-one fraction of the initial mass?}

Banks' construction is particularly useful because it illustrates why information storage and retention alone do not guarantee feedback. Information may be stored for a long time without affecting Hawking evaporation or the underlying quantum state transition. In order to achieve a genuine memory burden, the stored information must be written by the evolution, retained for a sufficiently long interval, and coupled back to subsequent evaporation kinetics. We refer to these three logically distinct requirements as \textsc{write}, \textsc{retain}, and \textsc{feedback}. 

The rest of the paper is structured as follows. In Sec.(\ref{MMO}), we study the conditions under which Hawking evaporation can be sufficiently affected by the correlations among the emitted radiation, assuming that an elementary step of the evaporation process deposits a dimensionless increment $\delta_k$ into a history-dependent variable. These are divided into three cases: short-range correlations, long-range correlations, and secular drift. In Sec.(\ref{retention}) we study the conditions under which the memory can accumulate coherently and be retained, and provide a no-go proposition, from which we identify conditions that may be broken to achieve memory burden. Afterward, in Sec.(\ref{sec:WRF}), we further explain how each stage of the  \textsc{write}-\textsc{retain}-\textsc{feedback} process of dynamical memory can be modeled. The explicit models can be found in Sec.(\ref{sec:hysteresis}), in which we show that memory-burdened black holes can have either a prolonged lifetime, or an infinite lifetime. In Sec.(\ref{sec:hysteresis2}) we show how, when accretion of matter is included, memory-burden effect can lead to hysteresis, which renders the mass-luminosity relation multivalued, meaning that we cannot infer the black hole mass just from its luminosity. This would have astrophysical implications, which we will discuss along with other issues, in the Discussion section, Sec.(\ref{Discussion}). We conclude in Sec.(\ref{Conclusion}).

\section{Macroscopic-Memory Requirement}
\label{MMO}

Let us start with the usual picture of Hawking evaporation.
For a Schwarzschild black hole, the averaged characteristic energy of a Hawking particle scales as
\begin{equation}
  \langle\omega\rangle=\frac{b}{M},
  \label{eq:omega}
\end{equation}
where $b$ is a dimensionless number depending on the emitted species and greybody factor.

The number emission rate scales as
\begin{equation}
  \Gamma_H \equiv \frac{\d N}{\d t} = \frac{\d N}{\d M}\frac{\d M}{\d t} =\frac{\d N}{\d M}\left(-\frac{\alpha}{M^2}\right).
  \label{eq:numberrate}
\end{equation}
On the other hand, we also have
\begin{equation}
  \Gamma_H\sim\frac{L_H}{\langle\omega\rangle}
	=\frac{M}{b}\frac{\alpha}{M^2}
  =\frac{\alpha}{bM}.
\end{equation}
Combining these two equations gives the rate of change of the mass as a function of particle number as
\begin{equation}
 \frac{\d M}{\d N} \simeq -\frac{b}{M}.
\end{equation}
Integration from the initial mass\footnote{Hereinafter we shall use subscript 0 to denote the initial value of a given quantity.} $M_0$ to some later mass $M$ gives the particle number as a function of the mass, roughly as
\begin{equation}
  N(M)\simeq\frac{M_0^2-M^2}{2b}.
  \label{eq:NofM}
\end{equation}
Since $S_\text{BH}=4\pi M^2$,
\begin{equation}
  N(M)=\mathcal{O}\left(S_0-S(M)\right).
\end{equation}
This is expected, since a rough estimate is usually that a Hawking photon has $\mathcal{O}(1)$ entropy, and the total emitted particle number is of the order of the Bekenstein-Hawking entropy.
Thus an order-one fractional change of the mass:
\begin{equation}
  M=xM_0,\qquad 0<x<1,\qquad x=\Ord(1),
\end{equation}
requires
\begin{equation}
  N(xM_0)=\Ord(S_0)
  \label{eq:NS}
\end{equation}
amount of entropy. This is well-known. The question for us is \emph{what happens when there is a memory-burden effect}?

Now let us suppose that the $k$th elementary step of the evaporation process deposits a dimensionless increment $\delta_k$ into a history-dependent variable,
\begin{equation}
  X_N=\sum_{k=1}^{N}\delta_k.
  \label{eq:sum}
\end{equation}
This incremental description specifically only concerns memory accumulated \emph{during evaporation}; it does not exclude memory patterns established earlier (e.g., during gravitational collapse), whose dynamical consequences may emerge only later.

We thus refer to an increment written by one elementary evaporation step as a ``deposit''.
The quantity $X_N$ is intentionally abstract as we do not know what it is, and it could be model-independent. For example, it may represent a coarse-grained memory load, a perturbation of a background variable, or any other quantity whose order-one growth can subsequently feed back on the evaporation law. In particular it should not automatically be identified with the fractional luminosity correction. Such an identification would require an additional feedback relation such as $\Delta L/L_H \propto X_N$ for small $X_N$.

Let us decompose each increment into a mean and a zero-mean fluctuation,
\begin{equation}
  \delta_k=\mu_k+\xi_k,
  \qquad
  \langle\xi_k\rangle=0.
  \label{eq:decompose}
\end{equation}
For now we will focus on the variable $\xi_k$ and return to the effect of the mean term $\mu_k$ later.

Over a portion of the evaporation during which the macroscopic parameters change only adiabatically, the connected correlation function may be treated locally as
\begin{equation}
  C(r)\equiv
  \langle\xi_k\xi_{k+r}\rangle.
  \label{eq:correlation}
\end{equation}

The variance of the accumulated stochastic part follows directly from Eq.\eqref{eq:sum} as:
\begin{equation}
  {\rm Var}(X_N)
  =
  \left\langle
  \left(\sum_{k=1}^{N}\xi_k\right)
  \left(\sum_{\ell=1}^{N}\xi_\ell\right)
  \right\rangle
  =
  \sum_{k,\ell=1}^{N}
  \langle\xi_k\xi_\ell\rangle.
  \label{eq:vardouble}
\end{equation}
There are $N$ diagonal terms with $k=\ell$, each of which equals $C(0)$.
For every separation $r>0$, there are $N-r$ pairs with $\ell=k+r$, together with the same number in the reversed ordering.
This gives the explicit Variance equation:
\begin{equation}
  {\rm Var}(X_N) = NC(0) + 2\sum_{r=1}^{N-1}(N-r)C(r).
  \label{eq:varianceexact}
\end{equation}
We shall now consider three different scenarios.

\subsection*{Case I: Short-Range Correlations}
Firstly, assume that the elementary fluctuations have root-mean-square (rms) amplitude
\begin{equation}
  \sigma\equiv \sqrt{C(0)}=\Ord(S_0^{-p})\Longrightarrow 
  C(0)=\sigma^2=\Ord(S_0^{-2p}).
  \label{eq:C0}
\end{equation}
Next, suppose the total correlation with neighboring emissions remains of the same parametric order as $S_0\to\infty$, i.e.,
\begin{equation}
  \sum_{r=0}^{\infty}|C(r)|=\Ord(S_0^{-2p}).
  \label{eq:summable}
\end{equation}
Physically, this means that each emission remains appreciably correlated with only an $\Ord(1)$ range of neighboring emissions in the macroscopic limit.
Using Eq.\eqref{eq:varianceexact}, we see that the variance is bounded from above:
\begin{equation}
  {\rm Var}(X_N)\leqslant 2N\sum_{r=0}^{\infty}|C(r)|=\Ord(NS_0^{-2p}).
\end{equation}
At a fixed mass fraction, $N=\Ord(S_0)$ by Eq.\eqref{eq:NS},  so $\Ord(N S_0^{-2p}) = \Ord(S_0^{1-2p})$.

We now denote
\begin{equation}
  X_{\rm rms}\equiv\sqrt{{\rm Var}(X_N)}.
  \label{eq:Xrmsdef}
\end{equation}
The results above then give
\begin{equation}
  X_{\rm rms}=\Ord\left(S_0^{1/2-p}\right).
  \label{eq:shortRangeScaling}
\end{equation}
For the natural case of an entropy-suppressed elementary fluctuation,
\begin{equation}
  |\delta_k|_{\rm rms}\sim S_0^{-1},
\end{equation}
i.e. $p=1$, and so one obtains
\begin{equation}
  X_{\rm rms} =\Ord(S_0^{-1/2}) \longrightarrow 0
  \label{eq:randomwalk}
\end{equation}
in the large black hole limit. For a solar mass black hole, in our units its initial entropy would be $S_0=\Ord(10^{77})$, so $X_{\rm rms}$ is about $10^{-39}$. For actual astrophysical black holes, whose masses are often much larger than a solar mass, this is practically zero.

This result means that total independent, correlation-free, emission is \emph{not} required for a macroscopic suppression. Any zero-mean sequence of $1/S_0$ deposits whose integrated correlation range stays finite exhibits the same parametric random-walk suppression.

It is worth noting that correlations between successive Hawking emissions can arise already from energy conservation and self-gravitation in the Parikh-Wilczek tunneling descriptions \cite{ParikhWilczek2000}, in which the probability of a later emission is conditioned on the energy previously radiated \cite{0903.0893}. However, in such a model, despite the fact that the next Hawking emission depends on the past, this dependence is only via the instantaneous current mass.
As such these correlations do not by themselves imply a macroscopic memory effect.

\subsection*{Case II: Long-Range Correlations}
 We now consider the case in which correlations persist across an increasing number of evaporation events\footnote{Entropy-dependent coherence scales have also been investigated previously in the contexts of black hole radiation, e.g. by Brustein and Medved \cite{1310.5861,1312.0880}, although their coherence number is not directly identifiable with the covariance of the memory increments considered in this work.}. Let us first quantify this physical range by defining a useful effective correlation length in the Hawking emission\footnote{Mathematically, this quantity is the ``integrated autocorrelation time'' familiar from statistical mechanics and correlated stochastic processes; see e.g., Eq.(12) of \cite{2107.06858}. However, the application here is as an effective number of correlated evaporation steps, rather than as a physical time or spatial length.}:
\begin{equation}
  \ell_{\rm corr} \equiv \frac{1}{C(0)}\left[ C(0)+2\sum_{r=1}^{\infty}C(r) \right].
  \label{eq:ellcorr0}
\end{equation}
A more useful version is the finite-$N$ truncation, but with a different weighting factor:
\begin{equation}
  \ell_{\rm corr}^{(N)} \equiv 1 + 2\sum_{r=1}^{N-1}\left(1-\frac{r}{N}\right)\frac{C(r)}{C(0)}.
  \label{eq:ellcorr}
\end{equation}
It is worth emphasizing that this definition does not introduce a new microscopic effect. It is merely a different description of the correlation function already present in Eq.\eqref{eq:varianceexact}.

But now suppose that under a specific model, the correlation length scales with the macroscopic entropy as
\begin{equation}
  \ell_{\rm corr}=\Ord(S_0^\nu),
  \qquad 0\leqslant \nu\leqslant 1.
  \label{eq:nu}
\end{equation}
This would lead to\footnote{The first equality is exact; the approximated version uses $\ell_{\rm corr}$, which may be mathematically more familiar.}
\begin{equation}
  {\rm Var}(X_N) = N C(0)\ell_{\rm corr}^{(N)} \simeq N C(0)\ell_{\rm corr},
\end{equation}
and therefore
\begin{equation}
  X_{\rm rms} =\Ord\left(S_0^{(1+\nu)/2-p}\right).
  \label{eq:masterScaling}
\end{equation}
A non-vanishing stochastic memory burden in the macroscopic limit requires $X_{\rm rms}=\Ord(1)$, and so
\begin{equation}
p \leqslant \frac{1+\nu}{2}.
\end{equation}
For $p=1=v$,  $X_{\rm rms}=\Ord(1)$ means $ \ell_{\rm corr} = \Ord(S_0)$. 

We can also heuristically write
\begin{equation}
  |X_N|\sim N^\beta|\delta|.
  \label{eq:beta}
\end{equation}
If $\ell_{\rm corr}\sim N^\nu$, by comparing with Eq.\eqref{eq:masterScaling}, we see that
\begin{equation}
  \beta=\frac{1+\nu}{2}.
\end{equation}
Ordinary diffusion corresponds to $\nu=0$ and $\beta=1/2$, while a fully coherent accumulation corresponds to $\nu=1$ and $\beta=1$.

The same scaling has an intuitive block interpretation.
If emissions are coherent over groups of roughly $\ell_{\rm corr}$ events, a correlated block contributes $\Delta X_{\rm block}\sim\ell_{\rm corr}\delta$, while there are approximately $N/\ell_{\rm corr}$ independent blocks.
Their random-walk sum therefore gives
\begin{equation}
  X_{\rm rms}
  \sim
  \sqrt{\frac{N}{\ell_{\rm corr}}}
  \left(\ell_{\rm corr}\delta\right)
  =
  \sqrt{N\ell_{\rm corr}}\,\delta.
  \label{eq:blockintuition}
\end{equation}
Thus $\ell_{\rm corr}=\Ord(1)$ reproduces ordinary diffusion, whereas $\ell_{\rm corr}=\Ord(N)$ approaches fully coherent accumulation.

\subsection*{Case III: Secular Drift}

Recall that in Eq.(\ref{eq:decompose}) we decomposed the increment $\delta_k$ into a mean part and a zero-mean part. In the previous two cases we exclusively considered the mean-free case. Now let us consider the effect of the $\mu_k$ terms. 
If, say,
\begin{equation}
  \mu_k=\frac{s}{S_0^p}
\end{equation}
with a \emph{common sign}, then 
\begin{equation}
  \langle X_N\rangle =s\frac{N}{S_0^p}=\Ord(S_0^{1-p})
\end{equation}
at fixed fractional mass loss, for which $N = \Ord(S^0)$.

In particular, for $p=1$,
\begin{equation}
  \langle X_N\rangle=\Ord(1),
\end{equation}
without any long-range stochastic correlations. 

We shall use the technical term \emph{secular}\footnote{Secular term in physics and mathematics refers to a component in an approximate series solution (such as in perturbation theory) that grows unbounded with time or position, breaking down the validity of the approximation over long periods. This is rather fitting for the Latin root of the word \emph{saecularis} (``of an age; occurring once in an age'').} for such a contribution when it accumulates systematically over many evaporation steps rather than averaging away (which is what happens if we do not impose the common-sign condition).

We have therefore established that both both secular drift and macroscopically correlated fluctuations can lead to macroscopic memory.

\section{Retention and NO-GO Result}\label{retention}

The preceding section asked how many elementary contributions can possibly accumulate coherently. The next important question for the memory-burden proposal is how long a contribution remains dynamically available before it is erased. While information erasure may sound alarming, we do not mean something as drastic as a violation of unitarity, just the milder version that the information becomes effectively unavailable to the black hole. That is, information is not lost, just inaccessible.

To make this relation explicit, suppose that an evaporation event writes an increment
\begin{equation}
  \delta_k=\mu+\xi_k,
  \qquad
  \langle\xi_k\rangle=0,
  \qquad
  \langle\xi_k^2\rangle=\sigma^2,
  \label{eq:retdecompose}
\end{equation}
into the memory variable. We remind the readers that ``deposit'' simply means the increment written by one elementary evaporation step. An unbiased deposit has $\mu=0$, whereas a biased deposit has a nonzero mean $\mu$, with a persistent sign, and therefore produces systematic drift. 

For a minimal solvable relaxation model, let each previously written contribution ``lose'' a fixed fraction per emission event.  Equivalently, the discrete update may be written schematically as
\begin{equation}
  X_{N+1}=e^{-\frac{1}{n_{\rm rel}}}X_N+\delta_{N+1},
\end{equation}
which is the event-count analogue of ordinary linear relaxation, $\dot X=-X/\tau_{\rm rel}+\text{source}$.  
The subscript ``rel'' is short for relaxation.
Iterating the recursive relation gives an exponential forgetting over $n_{\rm rel}$ emission events,
\begin{equation}
  X_N=\sum_{k=1}^{N} e^{-\frac{N-k}{n_{\rm rel}}}\delta_k.
  \label{eq:forgetting}
\end{equation}
For statistically independent zero-mean fluctuations, the variance is the geometric sum
\begin{equation}
  {\rm Var}(X_N)
  =
  \sigma^2\sum_{j=0}^{N-1}e^{-\frac{2j}{n_{\rm rel}}}
  =
  \sigma^2\left(
  \frac{1-e^{-\frac{2N}{n_{\rm rel}}}}
       {1-e^{-\frac{2}{n_{\rm rel}}}}\right).
  \label{eq:retvar}
\end{equation}
Likewise the systematic part gives 
\begin{equation}
  \langle X_N\rangle
  =
  \mu\sum_{j=0}^{N-1}e^{-\frac{j}{n_{\rm rel}}}
  =
  \mu\left(
  \frac{1-e^{-\frac{N}{n_{\rm rel}}}}
       {1-e^{-\frac{1}{n_{\rm rel}}}}\right).
  \label{eq:retmean}
\end{equation}

These exact expressions show explicitly how the elementary amplitude introduced in Eq.\eqref{eq:C0} enters the retained cumulative memory.
The exponential kernel discussed above is chosen for convenience and easy calculations to illustrate the underlying concept. For a general linear retention kernel $\mathfrak{K}$,
\[
  X_N=\sum_{k=1}^{N}\mathfrak{K}_{N-k}\delta_k,
\]
we can repeat the calculations to find that 
independent zero-mean deposits give ${\rm Var}(X_N)=\sigma^2\sum_j \mathfrak{K}_j^2$, whereas a secular mean gives $\langle X_N\rangle=\mu\sum_j \mathfrak{K}_j$. Any ordinary short-range kernel with effective width $n_{\rm rel}$ therefore yields, essentially, $X_{\rm rms}\sim\sigma\sqrt{n_{\rm rel}}$ and $\langle X_N\rangle\sim\mu n_{\rm rel}$ up to order-one factors. 

When $1\ll n_{\rm rel}\ll N$, using the appropriate series approximation, one finds that Eqs.\eqref{eq:retvar}--\eqref{eq:retmean} reduce to
\begin{equation}
  X_{\rm rms}
  \simeq
  \sigma\sqrt{\frac{n_{\rm rel}}{2}},
  \label{eq:relaxnoise}
\end{equation}
and
\begin{equation}
  \langle X_N\rangle
  \simeq
  \mu n_{\rm rel}.
  \label{eq:relaxdrift}
\end{equation}
More generally, for $n_{\rm rel}\gtrsim1$, the number of effectively remembered emissions is
\begin{equation}
  n_{\rm eff}\sim\min(N,n_{\rm rel}),
  \label{eq:neff}
\end{equation}
up to factors of order unity.

For the amplitudes $\sigma\sim S_0^{-p}$ and $\mu\sim S_0^{-p}$, one therefore has 
\begin{equation}
  X_{\rm rms}\sim S_0^{-p}\sqrt{n_{\rm eff}},
  \qquad
  \langle X_N\rangle\sim S_0^{-p}n_{\rm eff}.
  \label{eq:retainedscaling}
\end{equation}
Recall that at a fixed order-one mass fraction, $N\sim S_0$.  Hence for $p=1$, a systematically written memory can reach order unity only if
\begin{equation}
  n_{\rm eff}\sim S_0,
  \label{eq:nrelS}
\end{equation}
which in particular requires that the relaxation length not be parametrically shorter than $S_0$ emission events. By contrast, unbiased independent noise with even $n_{\rm rel}\sim S_0$ gives only $X_{\rm rms}=\Ord(S_0^{-1/2})$. To get an order-one stochastic effect still requires long-range correlations or another amplification mechanism.

We shall now translate an event-count retention length into a physical timescale.  Recall again that the characteristic Hawking number flux satisfies
\begin{equation}
  \Gamma_H \sim \frac{L_H}{\langle\omega\rangle} \sim \frac{M^{-2}}{M^{-1}} \sim M^{-1},
  \label{eq:numberfluxret}
\end{equation}
so the characteristic interval between any two emissions is approximately
\begin{equation}
  \Delta t_{\rm em}\sim\Gamma_H^{-1}\sim M.
  \label{eq:emissionintervalret}
\end{equation}
A memory extending over $n_{\rm rel}$ events therefore corresponds to the physical relaxation time
\begin{equation}
  \tau_{\rm rel} \sim n_{\rm rel}\Delta t_{\rm em}.
  \label{eq:taureldef}
\end{equation}
At a fixed mass fraction, $M\sim M_0$, so Eq.\eqref{eq:nrelS} gives
\begin{equation}
  \tau_{\rm rel} \sim S_0 M_0 \sim M_0^3 \sim t_{\rm life},
  \label{eq:slowmode}
\end{equation}
the lifetime of the black hole.
Likewise, the stochastic condition $\ell_{\rm corr}=\Ord(S_0)$ implies
\begin{equation}
  \tau_{\rm corr} \sim \ell_{\rm corr}\Delta t_{\rm em}=\Ord(M_0^3).
  \label{eq:corrtime}
\end{equation}
Therefore a macroscopic memory requires information to remain coherently relevant over the duration of a black hole lifetime, unless some other effects amplify it.

Before we move on to the no-go result, a few side remarks are in order.
The first concerns the discrete-time formulation above. This is natural given that Hawking evaporation is sparse \cite{1506.03975}, i.e., one particle is produced at random interval after another has been emitted. Nevertheless, many models of Hawking evaporation, including our subsequent concrete models, employ a continuous picture that is mathematically more convenient. Fortunately the scaling properties we discussed thus far do not depend on treating the radiation as a discrete sequence of classical particles. A continuous-time formulation is readily constructed as follows.

Let a history-writing source $\xi(t)$ feed a coarse-grained memory variable through a retarded kernel, schematically,
\begin{equation}
  X(t)=\int_0^t \d t'\,G(t,t')\,\xi(t').
  \label{eq:kernel}
\end{equation}
Then the mean is
\begin{equation}
  \langle X(t)\rangle=\int_0^t \d t'\,G(t,t')\langle\xi(t')\rangle,
  \label{eq:kernelmean}
\end{equation}
and the variance is
\begin{equation}
  {\rm Var}\,(X(t))=
  \int_0^t \d t'\int_0^t \d t''\,G(t,t')G(t,t'')
  C(t',t''),
  \label{eq:kernelvar}
\end{equation}
where
\begin{equation}
  C(t',t'')=\langle\xi(t')\xi(t'')\rangle_c
\end{equation}
is the continuous version of the correlation function.

Equations~\eqref{eq:kernelmean}--\eqref{eq:kernelvar} make the underlying statement more general:
an order-one memory requires either (1) a systematic source; (2) a response kernel with evaporation-scale support; (3) a correlation function with evaporation-scale support; or (4) dynamical amplification sufficient to compensate for their absence.

The next remark concerns \emph{when} the stochastic effects become appreciable. One may suspect that they are negligible all the way to the Planck regime since such effects need time to build up. However, other stochastic effects, such as the physical position drifting of a black hole due to the momentum recoil of Hawking emission events, already suggest that the effect can indeed appear much earlier. For the position drift, it amounts to $M_0^2$ change in the position in the evaporation timescale $M_0^3$ \cite{Page1980,1210.6348}.  
Indeed, the evaporation dynamics itself can amplify perturbations. To see this, 
we can linearize the mass loss rate
\begin{equation}
  \dot M=-\frac{\alpha}{M^2}
\end{equation}
around a background trajectory. This gives
\begin{equation}
  \delta\dot M=\frac{2\alpha}{M^3}\delta M.
\end{equation}
Using $\dot M=-\alpha/M^2$ again, the chain rule yields 
\begin{equation}
  \frac{\d(\delta M)}{\d M}=\frac{\d(\delta M)}{\d t}\left(\frac{\d M}{\d t}\right)^{-1}=
	-\frac{2}{M}\delta M.
\end{equation}
Upon integrating from the initial mass $M_i$ to the subsequent mass $M$, we observe that
\begin{equation}
  \frac{\delta M(M)}{\delta M(M_i)}=
  \left(\frac{M_i}{M}\right)^2.
\label{eq:amplification}
\end{equation}
Because $M < M_i$, this ratio is larger than unity.
Evaporation thus amplifies pre-existing mass perturbations, and the memory-burden effect can become appreciable while the black hole is still macroscopic. This we shall see explicitly when we have a concrete toy model later.

For now, we note that a systematic stochastic-gravity treatment by Hu and Roura \cite{HuRoura2007a,HuRoura2007b} had also found that induced fluctuations can grow and become important well before the Planckian regime. In their spherically symmetric analysis, under some assumptions, the fluctuations become comparable to the background when 
\begin{equation}
  M\sim M_0^{2/3}
\label{eq:M23}
\end{equation}
This example actually motivates a useful distinction:
\begin{align}
  &\text{(I) Pre-Planckian onset:} \nonumber\\ 
  & M/\mpl\rightarrow\infty,
  \\\nonumber
	\\\nonumber
  &\text{(II) Order-one-fraction onset:}  \nonumber\\
  & M/M_0\rightarrow x_\infty>0
  \quad\text{as }M_0/\mpl\rightarrow\infty.
\end{align}
Equation~\eqref{eq:M23} satisfies the first condition but not the second, since
\begin{equation}
  \frac{M}{M_0}\sim M_0^{-1/3}\rightarrow0
\end{equation}
in the large black hole limit.
Stochastic amplification therefore demonstrates that the naive random-walk estimate can be overcome dynamically, but it does not by itself provide an order-one-fraction memory-burden correction to Hawking evaporation.

We are finally ready to summarize the discussions thus far into a general result.
\begin{proposition*}{\textup{\textbf{[Conditional No-Go Proposition for
Macroscopic Memory Accumulation]}}}

Consider a family of macroscopic Schwarzschild black holes in the limit $S_0\to\infty$, observed at a fixed mass fraction $x=M/M_0\in(0,1)$. Let a history variable $X_N$ be accumulated from $N=\Theta(S_0)$
elementary deposits with vanishing mean and variance $C(0)=O(S_0^{-2p})$. Assume that their finite-window
integrated correlation factor, defined in Eq.(25), satisfies
\begin{equation}
    \ell_{\rm corr}^{(N)}=O(S_0^\nu),
    \qquad 0\leqslant\nu \leqslant 1.
\end{equation}
In the absence of additional dynamical amplification, the variance of the accumulated history variable obeys
\begin{equation}
    \operatorname{Var}(X_N)
    =O\left(S_0^{1+\nu-2p}\right),
\end{equation}
and hence
\begin{equation}
    X_{\rm rms}
    =O\left(S_0^{(1+\nu)/2-p}\right).
\end{equation}
Consequently, if
\begin{equation}
    p>\frac{1+\nu}{2},
\end{equation}
the accumulated history variable vanishes in mean square in the macroscopic limit.

If, furthermore, the feedback onto the evaporation luminosity has uniformly bounded sensitivity, such that
\begin{equation}
    \left|\frac{L(M,X_N)}{L_H(M)}-1\right|
    \leqslant \mathfrak{B}|X_N|,
\end{equation}
where $\mathfrak{B}>0$ does not grow with $S_0$, the corresponding fractional luminosity correction also vanishes.

In particular, for elementary deposits of rms magnitude $O(S_0^{-1})$, with zero mean and an integrated correlation range that grows sub-linearly with the initial entropy\footnote{Equivalently, $\lim_{S_0\to\infty} \left({\ell_{\rm corr}^{(N)}}/{S_0}\right)=0$, which is to say, the effective number of correlated emission events grows more slowly than the total number of Hawking emissions in the macroscopic limit. Such an integrated correlation range is said to be subextensive.}, $\ell_{\rm corr}^{(N)}=o(S_0)$, an order-one modification of Hawking evaporation cannot arise from unamplified stochastic accumulation alone.

\end{proposition*}

As with many ``no-go'' results, this proposition does not constitute a general prohibition against memory burden. 
Its main role is to identify the conditions under which a particular mechanism of memory accumulation cannot produce a macroscopic effect.
In other words, a proposed order-one history-dependent modification of evaporation must evade it via some means. For example:

\begin{enumerate}
\item \textit{Systematic writing.}
The history variable receives a nonzero mean deposit whose sign does not average away over the evaporation history.

\item \textit{Evaporation-scale retention.}
A systematically written memory survives for $n_{\rm rel}=\Ord(S_0)$ emissions, equivalently for a time $\tau_{\rm mem}=\Ord(t_{\rm life})$.

\item \textit{Macroscopic correlations or collective constraints.}
Zero-mean microscopic deposits remain correlated across an $\Ord(S_0)$ range of emission number.

\item \textit{Secular instability or amplification.}
Initially small fluctuations are dynamically amplified strongly enough to overcome the accumulation bound before the desired mass scale.

\item \textit{Nonlocal or global feedback.}
Information stored in radiation or global correlations is returned to the local transition dynamics through a channel not captured by short-range, rapidly relaxing semiclassical response.

\item \textit{Dynamically efficient feedback.}  
The retained history variable must couple strongly enough to the evaporation law to produce an observable effect. Schematically,
\begin{equation}
\dot M=-L_H(M) \Phi(q),
\end{equation}
where $q$ denotes a memory variable in general\footnote{In the sections leading up to the No-Go Proposition, we have used $X_N$ as a generic accumulated history variable. It is deliberately abstract and indexed by emission number $N$, because the question we were interested in was: given microscopic deposits $\delta_k$, how large can the accumulated record become? By contrast, $q=q(t)$ is a coarse-grained dynamical memory variable (i.e. an effective state variable instead of a microscopic writing) that has its own time evolution, retention/relaxation, and feedback onto the mass loss rate $\dot {M}$. The simplest case would have $q \propto X_N$ but this is not a general requirement.}. 

\end{enumerate}

Note that items 1 to 5 mostly address how a record is written and survives, while item 6 makes explicit the final dynamical step. The model explored by Banks \cite{Banks2026}, for example, illustrates how a putative memory burden need not produce effective macroscopic feedback, since the penalty associated with frozen degrees of freedom can be offset by the larger phase space available to decay products.
Note that $q$ alone does not determine the size of the modification: it must be supplemented by a microscopic relation specifying $\Phi(q)$.  An order-one change may arise either because $q$ itself becomes macroscopic or because the feedback is sufficiently nonlinear or parametrically sensitive that a smaller retained memory produces an order-one change in the luminosity.
We also remark that items 1 to 5 are not mutually exclusive channels, and a successful theory or model may realize several at the same time. 

So far we have discussed how a model can realize memory burden in an evaporating black hole spacetime. We shall move on with concrete toy models to see how a memory-burden effect can actually affect the evaporation history of the black hole, but first we need to make clear how such models may be constructed.


\section{Modeling Dynamical Memory}
\label{sec:WRF}

Let us start by summarizing the main idea thus far. We have examined via basic stochastic language how memory burden can be realized while the black hole is still macroscopic. Crucially we have modeled the ``memory deposits'' at each emission as an increment $\delta_k$ that writes into a history-dependent variable $X_N$, as given by Eq.\eqref{eq:sum}. We then discuss the retention of the memory in Sec.(\ref{retention}) via the retention kernel $\mathcal{K}$, with the exponential kernel as an explicit example. Thus already we have seen that a memory burden necessitates the record of the memory, and the retention of the memory.
Indeed, information may be stored for a long time without affecting the evaporation of a black hole. In fact this is the standard scenario in which the information is believed to be encoded in the emitted radiation via nontrivial quantum entanglement. For the memory-burden scenario, if we want the stored memory to become a \emph{dynamical memory}, a record must have been written at some stage, retained for a sufficiently long duration, and coupled back to the subsequent evaporation process. The last step was not explicitly discussed thus far, but will become clear later.

To summarize, there are three distinct steps for the realization of memory-burden effect in Hawking evaporation:\\
\begin{center}
{\large
\vintageeq{
 \textgoth{Write}\;\longrightarrow\;
  \textgoth{Retain}\;\longrightarrow\;
  \textgoth{Feedback} \nonumber
  \label{eq:WRF}
	}
}
\end{center}

Let us now explain how we can model each of these steps.
\begin{itemize}
\item[(I)]\textbf{WRITE: The evolution must create a history variable.} In addition, such a variable should be able to evolve in time. We will model such a quantity $q$, which is a coarse-grained variable that is not necessarily entirely fixed by the instantaneous mass and other conserved charges of the black hole. It evolves via, schematically,
\begin{equation}
\dot{q} = \mathcal{W},
\end{equation}
where $\mathcal{W}$ is the ``history-writing'' source term that depends on the black hole states, including the mass. 

As mentioned before in Sec.(\ref{MMO}), the writing of a memory record need not occur exclusively during Hawking evaporation. A black hole may already possess a stored memory pattern established during its formation, whose energetic cost or dynamical influence changes as the black hole subsequently loses mass. Such a scenario need not involve continuously writing new information during evaporation. In this more general setting, the variable $q$ may represent the effective burden associated with a pre-existing record, rather than the amount of newly deposited information. The source $\mathcal{W}$ should then be understood as governing the evolution of this effective burden, which may include changes in the accessibility or energetic cost of previously stored information. In view of this, it is worth emphasizing that, in general, a pre-existing memory record and an evolving memory burden are not necessarily the same variable. For example, a fixed stored memory pattern $\mathcal{I}_0$ could produce an evolving $q=q(M,\mathcal{I}_0)$ as the black hole evaporates, even without any new information being written. 

\item[(II)]\textbf{RETAIN: The record must survive.} A written history variable may nevertheless become inaccessible to the black hole. Note that this does not require the information to be destroyed (which would be in tension with unitarity), merely that it is effectively relaxed or erased as far as the evaporating black hole is concerned. Thus, to the source term above we add a relaxation term $\mathcal{R}$ as follows:
\begin{equation}
\dot{q} = \mathcal{W}-\mathcal{R}[q],
\end{equation}
Near a regime in which a linear approximation is adequate, we can have a simple explicit model with
\begin{equation}
\mathcal{R}[q]=\gamma q,
\end{equation}
in which $\gamma^{-1}$ is the relaxation timescale.

\item[(III)]\textbf{FEEDBACK: Stored History Must Modify Subsequent Evolution.} 
To affect the evolution under Hawking evaporation, the luminosity must be a function of both the conserved charges as well as of the history variable $q$. For a Schwarzschild black hole, this means 
\begin{equation}
\dot{M} = -L(M,q).
\end{equation}
The standard memory-less Hawking evaporation corresponds to $L_H (M) \equiv L(M,0)$, while to satisfy the Feedback requirement simply means that
\begin{equation}
\frac{\partial L(M,q)}{\partial q} \neq 0,
\end{equation}
and hence the equation $\dot{M}$ together with $\dot{q}$, become a coupled ODE system.
\end{itemize}

In the preceding sections we have assumed the standard Hawking temperature and luminosity to hold. However, once we allow memory burden in general, this may no longer be true. Still, the simplest scenario of memory burden does not require any modification to the Hawking temperature; we only need to modify the rate at which the emissions occur. We will refer to this type of models as the \emph{kinetic memory} subclass.  
Bekenstein and Meisels showed that Hawking emission and absorption coefficients obey the usual degeneracy relation, with $e^{S_{\rm BH}}$ playing the role of the state degeneracy \cite{BekensteinMeisels1977}.
Similarly, energy-conserving tunneling treatments organize the emission probability through the change in black hole entropy. Schematically \cite{ParikhWilczek2000},
\begin{equation}
  \Gamma\propto e^{\Delta S_{\rm BH}}.
  \label{eq:tunneling}
\end{equation}

Detailed balance fixes a \emph{ratio} of forward and reverse rates but not their common kinetic prefactor $K_\text{kin}$.
Schematically one may write
\begin{equation}
\begin{cases}
  \Gamma_{\downarrow}(M,q)
  =K_\text{kin}(M,q)\,\Gamma_{\downarrow}^{(0)}(M),\\
  \Gamma_{\uparrow}(M,q)
  =K_\text{kin}(M,q)\,\Gamma_{\uparrow}^{(0)}(M),
\end{cases}
  \label{eq:kineticfactor}
\end{equation}
so that
\begin{equation}
  \frac{\Gamma_{\downarrow}}
       {\Gamma_{\uparrow}} =
  \frac{\Gamma_{\downarrow}^{(0)}}
       {\Gamma_{\uparrow}^{(0)}}
\end{equation}
even when $K$ depends on a non-equilibrium history variable.
``Kinetic'' refers to the dynamical transition rates: thermodynamic state counting fixes the relative weighting of forward and reverse transitions, whereas the common prefactor fixes how quickly either transition occurs.

This allows a realization of black hole kinetic aging. Two black holes with the same $M$, same entropy $S_{\rm BH}(M)$, and same equilibrium temperature $T_H(M)$ could in principle still have different luminosities because their history-dependent common prefactors differ.
Of course Eq.~\eqref{eq:kineticfactor} merely shows that such a dependence is not excluded by detailed balance alone, a specific microscopic model would still have to derive the explicit dependence $K_\text{kin}(M,q)$.

Another possibility is that in addition to the total luminosity, the distribution of the emitted particle energies is also modified. A characteristic emitted energy of order $1/M$ does not necessarily imply an exactly thermal spectrum. In fact in a genuinely nonequilibrium state there need not be a unique temperature that can be inferred from the radiation. Therefore it is conceivable that memory burden may change the normalization factor or the spectral shape of the spectrum. We refer to this broader possibility as \emph{spectral memory} but shall not pursue it any further (an example in which the spectrum was modified by memory burden can be found in \cite{2604.06858}). Nevertheless we wish to emphasize that, since the memory-burden mechanism is intended to signal a breakdown of Hawking semiclassical evaporation after the stored memory becomes dynamically important \cite{DvaliEtAl2020,DvaliEtAl2024}, spectral memory may play rather important roles in the late-time evaporation once the effect of memory burden becomes important. Of course, it is also possible that the thermodynamic temperature itself may be modified in some \emph{thermodynamic-memory} subclass, which we also do not explore further in this work. It suffices to mention that mass-loss rate alone cannot distinguish kinetic suppression from a reduced temperature. To distinguish the thermodynamic subclass from the kinetic one, we should compare their transition information.
For example, for the kinetic-memory subclass, one should have
\begin{equation}
  \frac{\Gamma_\text{emit}}
       {\Gamma_\text{absorb}} \sim e^{-\omega/T_H},
\end{equation}
while for the thermodynamic-memory subclass, since the detailed-balance temperature has been modified into $T_\text{mod}$, one should have
\begin{equation}
  \frac{\Gamma_\text{emit}}
       {\Gamma_\text{absorb}} \sim e^{-\omega/T_\text{mod}}.
\end{equation}

Suppression of the particle production rate without requiring that the Hawking temperature vanishes is not unique to our memory-based models. In the UV-modified evaporation scenarios of Ho et al. \cite{2309.01638,2606.20314}, Hawking emission can terminate around the scrambling time while the Hawking temperature remains approximately unchanged. In that case, however, the suppression in emission is due to trans-Planckian or nonlocal dynamics rather than from an accumulated memory degree of freedom.

\section{Evaporation with Memory Burden and Lifetime Analysis}
\label{sec:hysteresis}

Let us now consider a minimal phenomenological model of the kinetic-memory subclass:
\begin{align}
\begin{cases}
  \dot M &=-L_H(M)\Phi(q),\\
	\\
  \dot q &=\eta(M)L_{\rm w}-\gamma(M)q,
\end{cases}
  \label{eq:phenq}
\end{align}
with the conditions that $\Phi(0)=1$ and $\Phi'(q)\leq 0$ for $q\geqslant 0$ so that the first ODE describes a suppression of the luminosity. In order to have the same physical dimension as $\dot{q}$, we see that $\gamma$ cannot be dimensionless. In fact, $\gamma^{-1}$ should be interpreted as a memory-relaxation or erasure time scale. On the other hand, the quantity $L_{\rm w}$ is the luminosity associated with the writing of the memory, and in general need not equal the net mass-loss rate. We shall now examine a concrete model. We will assume that $q(0)=0$ (no initial effective burden from formation etc.) for simplicity.

Let us take a concrete feedback law given by the Hill form\footnote{This form of a function was first used in biochemistry. Specifically it appears in the Hill equation for quantifying the binding of oxygen to haemoglobin in 1910. The exponent $n$
is a parameter that expresses the degree of departure from the earlier model of Michaelis–Menten kinetics (corresponding to $n=1$). In Hill's paper \cite{Hill1910}, he explicitly said that his model is phenomenological and did not provide any direct physical meaning on the parameters (though this was provided by the follow-up works by many authors). We are using the Hill model in black hole evaporation much in the same spirit, namely, only as a phenomenological one.} \cite{Hill1910}
\begin{equation}
  \Phi(q)=\frac{1}{1+(q/q_c)^n},
  \label{eq:hill}
\end{equation}
with a threshold $q_c$ and a steepness index $n$. Inspired by biological enzymatic kinetic modeling, this function provides a smooth interpolation between an unsuppressed regime, $q\ll q_c$, and a memory-burdened regime, $q\gtrsim q_c$. This is of course just a practical choice to see how such a model may work\footnote{A mathematically similar rational-power suppression factor has also been employed in the memory-burdened PBH literature \cite{2604.06858}, although in that work it parametrizes an effective redshift suppression of the observed flux rather than the feedback from a dynamical memory variable.}, and is not meant to be realistic. A realistic memory-burden model should derive an explicit form of $\Phi(q)$, presumably from a microscopic theory, or otherwise motivated by some physical considerations. 

Now, define the dimensionless mass and time by normalizing them with respect to the standard Hawking lifetime:
\begin{equation}
  x\equiv\frac{M}{M_0},
  \qquad
  \tau\equiv\frac{t}{t_H},
  \qquad
  t_H\equiv\frac{M_0^3}{3\alpha}.
  \label{eq:dimensionlessHill}
\end{equation}
Namely, $t_H$ is the standard Schwarzschild lifetime in the geometric optic approximation $\dot{M}=-L_H=-\alpha/M^2$.
Furthermore, consider the simplest case in which the writing luminosity is identical to the emitted luminosity itself, $L_{\rm w}=L=-\dot M$, and take a constant writing efficiency $\eta=w/M_0$. 
Let us also define the dimensionless quantity $\lambda\equiv\gamma t_H$.
Eq.(\ref{eq:phenq}) then becomes the coupled ODE

\begin{align}
\begin{cases}
  \dfrac{\d x}{\d\tau}
  &=-\dfrac{\Phi(q)}{3x^2},
 \\
\\
  \dfrac{\d q}{\d\tau}
  &=-w\!\left(\dfrac{\d x}{\d\tau}\right)-\lambda q.
  \label{eq:hillqdot}
\end{cases}
\end{align}
The parameters now have a clear physical interpretation: $w$ is a dimensionless writing strength measuring the memory written per unit fractional mass loss, while $\lambda^{-1}$ is the retention time in units of the standard Hawking lifetime.

For the erasure-free case, $\lambda=0$, and the second equation integrates exactly to give
\begin{equation}
  q(\tau)=w[1-x(\tau)],
  \label{eq:qexactHill}
\end{equation}
for the initial condition $x(0)=1$ and $q(0)=0$.
The Hill model therefore reduces to a single separable differential equation,
\begin{equation}
  \frac{\d\tau}{\d x} =-3x^2\left\{1+\left[\frac{w(1-x)}{q_c}\right]^n\right\}.
\end{equation}
Its solution may be written as
\begin{equation}
  \tau(x)
  =1-x^3 +3\left(\frac{w}{q_c}\right)^n\int_x^1 \,u^2(1-u)^n \d u.
  \label{eq:hillquadrature}
\end{equation}
The integral is of the Chebyshev form and can be written in terms of the incomplete beta function\footnote{The incomplete beta function is defined by \begin{equation} B_z(1+a,1+b)\equiv \int_0^z y^a(1-y)^b \d y.\end{equation}} as
\begin{equation}
  \tau(x)
  =1-x^3 +3\left(\frac{w}{q_c}\right)^n
  \left[B(3,n+1)-B_x(3,n+1)\right].
  \label{eq:hillbeta}
\end{equation}
As a consistency check, let us note that the term $1-x^3$ recovers precisely the standard Schwarzschild result. The new term in the square bracket is the accumulated delay produced by memory feedback.

As a concrete illustration, take the parameter values to be
\begin{equation}
  n=4,\qquad q_c=1,\qquad w=3.
  \label{eq:hillnumbers}
\end{equation}
The feedback reaches half strength, $\Phi=1/2$, when $q=q_c$, for which the normalized mass is $x=2/3$.
By $x=1/2$, the original mass has been halved by the evaporation, and one has $q=3/2$ and $\Phi=16/97\simeq0.165$.

\begin{figure}[b]
  \centering
  \includegraphics[width=0.98\linewidth]{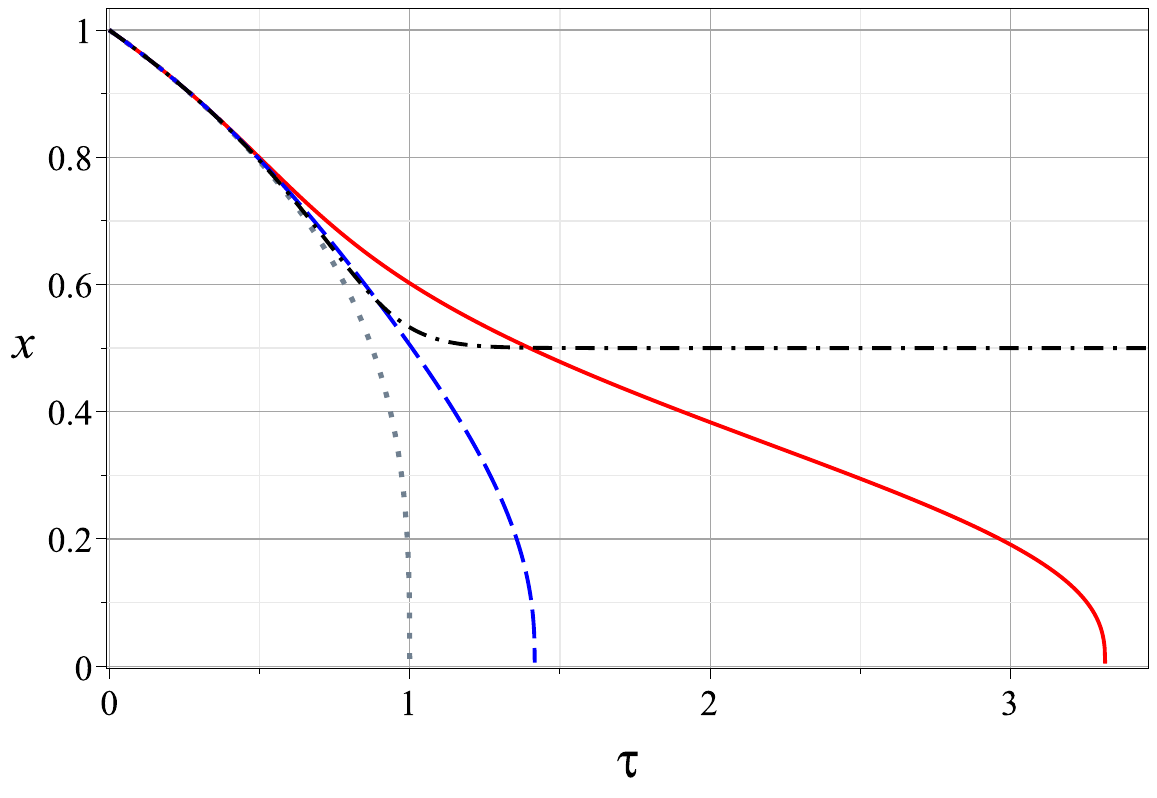}
  \caption{Four phenomenologically distinct evaporation histories with the vertical axis being the black hole mass normalized with its initial value: $x\equiv M/M_0$; and the horizontal axis is the normalized time $\tau\equiv t/t_H$, where $t_H$ is the Standard Hawking lifetime $t_H \sim M_0^3$. In dotted grey is the standard Schwarzschild evaporation that reaches zero mass at $\tau=1$. In solid red is the erasure-free Hill model with $n=4$, $q_c=1$ and $w=3$, exhibiting an order-one enhancement in the lifetime but still evaporates completely at $\tau=116/35\simeq3.31$. In dashed blue curve is the same Hill model but with the erasure term chosen as $\lambda =\gamma t_H =1$. Finally, the dash-dotted black curve is the example of Eq.\eqref{eq:terminalnumbers} that approaches the terminal state $M/M_0=1/2$ asymptotically because its kinetic factor vanishes. All modified curves are illustrative toy models rather than realistic physical models.}
  \label{fig:hillp1}
\end{figure}

The total lifetime follows directly from Eq.~\eqref{eq:hillbeta} by setting $x=0$:
\begin{equation}
  \frac{t_{\rm life}}{t_H}
  =1+3\left(\frac{w}{q_c}\right)^4 B(3,5)
  =1+\frac{81}{35}
  =\frac{116}{35}
  \simeq 3.31.
  \label{eq:hilllifetime}
\end{equation}
This is slightly more than 3 times the standard lifetime under the standard memory-less Hawking evaporation. Nevertheless, the lifetime is still finite. This is a concrete demonstration that memory burden can prolong the lifetime of an evaporating black hole while still allowing evaporation to complete in a finite time. We shall plot the mass evolution of this model, along with several others, in Fig.(\ref{fig:hillp1}) later.

It is of course possible to have a scenario in which the black hole stops evaporating. For a concrete example,
take again $q=w(1-x)$ but introduce a terminal mass $M_\star$ via
\begin{equation}
  q_\star=w(1-x_\star),
  \qquad x_\star\equiv\frac{M_\star}{M_0},
\end{equation}
and, for $q<q_\star$, instead of the preceding Hill form, choose the terminal feedback factor as
\begin{equation}
  \Phi_{\rm term}(q)
  = \left[1-\left(\frac{q}{q_\star}\right)^m\right]^r.
  \label{eq:terminalfeedback}
\end{equation}

Power-law functions of this general type\footnote{The form of Eq.(\ref{eq:terminalfeedback}) has been used in the literature in other contexts. Notably, in the statistical methods of hydrology, the Kumaraswamy distribution has cumulative distribution function of the form $F(x)=1-(1-x^a)^b$, and thus a survival function of the form $\bar{F}(x)=1-F(x)=(1-x^a)^b$. Similarly, the characteristic factor in the $q$-Weibull distribution function, after a suitable re-definition of variables, can also be written in such a form.} already appear in the memory-burden literature. In particular, Dvali's assisted-gaplessness construction describes memory-mode energy gaps proportional to
$(1-n_0/N_c)^n$, with both the linear case and higher order case considered \cite{1812.08749,2503.21740}. Our terminal feedback law is phenomenological,
with $m$ controlling the shape of the suppression away from the endpoint, whilst $r$ controls the order of vanishing at the endpoint. The exponent $m$ may be viewed as modeling a response to the growing memory-mode gap or the associated burden, but neither the precise feedback function nor its vanishing at a finite critical value follows from that microscopic Hamiltonian.

Taking specific values for our model with
\begin{equation}
  x_\star=\frac{1}{2}, \quad w=2,\quad q_\star=1,\quad m=4,\quad r=1,
  \label{eq:terminalnumbers}
\end{equation}
gives the terminal feedback factor written now as a function of the normalized mass:
\begin{equation}
  \Phi_{\rm term}(x)
  =1-[2(1-x)]^4.
  \label{eq:terminalphix}
\end{equation}
In this case, the lifetime is infinite. This can be seen by expanding the terminal feedback factor in a series around its endpoint $x=x_\star=1/2$:
\begin{equation}
\Phi_{\rm term}(x) = 8\left(x-\frac{1}{2}\right)+\mathcal{O}\left(\left(x-\frac{1}{2}\right)^2\right).
\end{equation}
Using the fact that the mass-loss rate is
\begin{equation}
\frac{\d x}{\d \tau}=-\frac{\Phi}{3x^2}, 
\end{equation}
one finds that the leading behavior is governed by
\begin{align}
\frac{\d \varepsilon}{\d \tau}&=-\frac{\Phi}{3x^2} = -\frac{8\varepsilon}{3}\left(\varepsilon + \frac{1}{2}\right)^{-2}+ \Ord(\varepsilon^2)\\
&=  -\frac{32\varepsilon}{3} + \Ord(\varepsilon^2), 
\end{align}
where we have defined $\varepsilon \equiv x-1/2 \ll 1$.
This yields, asymptotically, the exponential decay
\begin{equation}
x(\tau) \simeq \frac{1}{2}+ \text{const.}e^{-\frac{32}{3}\tau}, \qquad \tau \gg 1.
\end{equation}
We will show its plot in Fig.(\ref{fig:hillp1}) in comparison to other models.

In other words, the black hole mass asymptotically approaches $M_\star$, and therefore it becomes an effective remnant. However, crucially, unlike an extremal black hole, this remnant state has a finite nonzero temperature\footnote{This should be compared and contrasted with the models in \cite{0106080, 2505.07972}, in which the black hole stops evaporating at the Planck scale \emph{in a finite time}, but nevertheless has a nonzero temperature.}. This is because the mass loss rate is $\dot{M}=-L_H(M)\Phi(q)$, and so $L_H(M) \sim T_H^4$ can be nonzero but the factor $\Phi \to 0$. We shall refer to this scenario as \emph{kinetic freezing} of the black hole. Note that in this scenario, the actual emitted energy flux also goes to zero. The black hole is \emph{not} continuing to emit particles while its mass stays fixed. What remains finite is only the underlying thermodynamic scale $T_H=1/8\pi M$, not the physical luminosity. In the limiting frozen state, there may be essentially no outgoing quanta with which to measure that temperature operationally.

We now plot in Fig.\eqref{fig:hillp1} the mass evolution under the standard Hawking evaporation, under the Hill model with and without erasure term, and the terminal feedback model. 

Let us take a closer look at when a black hole in the kinetic-memory subclass can have an infinite lifetime. Suppose, as in the terminal feedback example above, that the memory variable is single-valued, $q=q(M)$, we define
\begin{equation}
  \Phi_{\rm eff}(M)\equiv\Phi[q(M)].
\end{equation}
The mass loss rate is
\begin{equation}
  \dot M=-\frac{\alpha}{M^2}\Phi_{\rm eff}(M),
  \qquad \Phi_{\rm eff}(M)\geqslant 0,
  \label{eq:generalkineticendpoint}
\end{equation}
and hence the elapsed time between the final terminal mass $M_f$ and the initial mass $M_i$ is
\begin{equation}
  t(M_f)-t(M_i)
  =
  \frac{1}{\alpha}
  \int_{M_f}^{M_i}
  \frac{M^2}{\Phi_{\rm eff}(M)}\,\d M.
  \label{eq:lifetimeintegral}
\end{equation}
The final state is obtained in a finite time if and only if the mass integral above converges.

Suppose first that the evaporation approaches a nonzero limiting mass $M_\star>0$ and near the endpoint 
\begin{equation}
  \Phi_{\rm eff}(M)\sim \mathcal{A}_{\star}(M-M_\star)^r,
  \qquad \mathcal{A}_{\star}>0.
  \label{eq:finiteendpointpower}
\end{equation}
Since the factor $M^2$ tends to the nonzero constant $M_\star^2$, Eq.\eqref{eq:lifetimeintegral} tells us that there will be a finite endpoint for $r < 1$, and infinite one if $r \geqslant 1$. 
The marginal case $r=1$ gives, for the mass gap $\Delta M\equiv M-M_\star$, 
\begin{equation}
\frac{\d (\Delta M)}{\d t}\simeq -\frac{\alpha \mathcal{A}_{\star}}{M_\star^2}\Delta M.
\end{equation}
Denote $\mathfrak{C}\equiv {\alpha \mathcal{A}_{\star}}/{M_\star^2} > 0$. Then the mass gap decays exponentially, or equivalently, 
\begin{equation}
M(t)=M_\star+\Delta M(t_0)e^{-\mathfrak{C} (t-t_0)}.
\end{equation}
For $r>1$, the mass gap decays as a power law instead.

What happens when the endpoint is at $M_\star=0$? Then 
\begin{equation}
  \Phi_{\rm eff}(M)\sim \mathcal{A}_0 M^{\delta_{\rm end}},  \qquad \mathcal{A}_{0}>0,
  \label{eq:zeroendpointpower}
\end{equation}
and thus a putative endpoint would satisfy
\begin{equation}
  t_{\rm end}\sim\int_{0}^{M_i} M^{2-\delta_{\rm end}}\,\d M.
\end{equation}
From this integral we can conclude that finite lifetime corresponds to $\delta_{\rm end} < 3$, while infinite lifetime corresponds to  $\delta_{\rm end} \geqslant 3$.

Interestingly an independent consistency check exists for the zero-mass criterion.
For a Schwarzschild black hole with a \emph{negative} GUP (generalized uncertainty principle) parameter, Ref.\cite{Ong2018ZeroMass} found that in the small mass regime, the mass loss rate scales as
\begin{equation}
  \dot{M}
  \sim -C_{\rm GUP}M^2, \qquad C_{\rm GUP}>0,
  \label{eq:gupsmallM}
\end{equation}
and the mass decays as $M\propto t^{-1}$, thus approaches zero asymptotically\footnote{Unlike the positive GUP parameter case which is associated with a minimal length or minimal mass, the negative GUP parameter does not admit a minimal mass or length scale. However because the evaporation takes infinitely long time, there is still an effective black hole remnant \cite{Ong2018ZeroMass}.}. 
Comparing Eq.\eqref{eq:gupsmallM} with Eq.\eqref{eq:generalkineticendpoint} shows that the GUP correction is equivalent, at the level of late time mass loss scaling, to
\begin{equation}
  \Phi_{\rm GUP}(M)\propto M^4.
\end{equation}
and so 
$\delta_{\rm end}^{\rm GUP}=4$,
which agrees with our infinite lifetime criterion ($\delta_{\rm end} \geqslant 3$) above.
This provides a non-memory example of the same endpoint test and illustrates the broader point that a finite nonzero limiting temperature does not by itself determine whether an evaporation endpoint is reached in a finite time \cite{Ong2018ZeroMass,YaoHouOng2019}.

We can comment a bit more about the kinetic freezing scenario. As we have just seen in the example above, this is only approached asymptotically. In fact, this is a generic phenomenon. 

Consider the system specified by
\begin{equation}
  \dot M=-F(M),~
  F(M)>0\ \text{for }M>M_\star,
  ~
  F(M_\star)=0.
  \label{eq:onedimF}
\end{equation}
Local Lipschitz continuity implies that sufficiently near the terminal mass $M_\star$,
\begin{equation}
  0\leqslant  F(M)\leqslant K|M-M_\star|
  \label{eq:LipschitzF}
\end{equation}
for some finite constant $K$.
With $\Delta M=M-M_\star>0$,
\begin{equation}
  \dot{\Delta M}=-F(M)\geqslant -K\Delta M,
\end{equation}
so multiplying $\dot{\Delta M}+K\Delta M$ by the integrating factor\footnote{Alternatively, apply (the minus sign version of) Grönwall's inequality.} gives 
\begin{equation}
  \frac{\text{d}}{\text{d} t}\!\left(e^{Kt}\Delta M\right)
  =e^{Kt}\left(\dot{\Delta M}+K\Delta M\right)
  \geqslant 0.
\end{equation}
Integrating from $t_0$ to $t$ implies that, for every finite $t$,
\begin{equation}
  \Delta M(t) \geqslant
  \Delta M(t_0)e^{-K(t-t_0)} >0.
  \label{eq:Lipschitzendpoint}
\end{equation}

Thus under the Lipschitz continuity assumption, a finite mass gap can be closed no faster than the exponential decay. Consequently a smooth or analytic kinetic factor that truly vanishes at a finite-mass fixed point must produce an asymptotic state rather than a finite-time endpoint. In fact we can give a more general formulation of this ``unattainability theorem'' beyond the one-dimensional kinetic subclass.

Suppose the effective coarse-grained dynamics can  be described by an autonomous system
\begin{equation}
  \dot{\bm y}=\bm F(\bm y),
  \qquad
  \bm y=(M,q,\ldots),
  \label{eq:dynsystem}
\end{equation}
and let $\bm y_\star$ be a genuinely frozen configuration,
\begin{equation}
  \bm F(\bm y_\star)=0.
  \label{eq:fixedpoint}
\end{equation}
Here ``genuinely frozen'' means that all coarse-grained state variables have stopped evolving, $\dot M=\dot q=\cdots=0$. In contrast, a moment at which only $\dot M=0$ while another variable continues to evolve is not a fixed point of the full system. 

The special role of $\bm y_\star$ is precisely that Eq.\eqref{eq:fixedpoint} makes the constant curve $\bm y(t)=\bm y_\star$ a solution of the equations of motion.
If $\bm F$ is locally Lipschitz near $\bm y_\star$, the initial-value problem with data $\bm y(t_\star)=\bm y_\star$ has a unique local solution, namely the aforementioned constant solution.
Suppose instead that a genuinely evolving trajectory reached $\bm y_\star$ at some finite time $t_\star$.
At $t_\star$ the evolving trajectory and the constant solution would solve the same initial-value problem, and uniqueness would force them to coincide locally, contradicting the assumption that the incoming trajectory was distinct and evolving immediately before arrival.
Note that this argument is special to a fixed point.  At an ordinary point $\bm y_0$ with $\bm F(\bm y_0)\neq0$, the constant curve $\bm y(t)=\bm y_0$ is not a solution, and there is no obstruction to the unique moving trajectory passing through $\bm y_0$ at finite time. Hence a regular fixed-point freezing is unattainable in finite time. If the memory is not genuinely frozen, say, if $q_\star > 0$, and a positive erasure rate remains active, then $\dot{q}=-\lambda q_\star \neq 0$ at the putative endpoint. The memory would then be able to relax and the suppression could lift. 

The locally Lipschitz condition is crucial for this result to hold.
The elementary contrasting example is $\dot x=-x$, whose Lipschitz fixed point at $x=0$ is approached exponentially but never reached at finite time, versus $\dot x=-\sqrt{x}$ for $x\geqslant 0$, which reaches $x=0$ in a finite time precisely because the Lipschitz condition is violated at $x=0$. This unattainability result can be compared with the so-called ``complementary third law'' of Ref.\cite{YaoHouOng2019}, which states that under some regularity assumptions, a black hole cannot evaporate to a massless finite nonzero temperature end state in a finite time.

The aforementioned loophole involving regularity conditions is important. Finite-time halting of the evaporation \emph{is} possible if the effective evolution is non-Lipschitz at the endpoint, as in the $0<r<1$ branch of Eq.(\ref{eq:finiteendpointpower}), or if an absorbing boundary is imposed by hand, among other possibilities.  An accretion or rewriting process may also change the accessible state space.
The unattainability statement therefore applies to a regular fixed-point freezing, not to every conceivable prescription that sets the mass-loss rate to zero.

Now that we have mentioned accretion, this is a natural point to move on to the next important topic: hysteresis.

\section{HYSTERESIS Due to memory burden}
\label{sec:hysteresis2}

Although the non-equivalence of old and young black holes at the same mass has already been emphasized in the foundational memory-burden literature \cite{DvaliEtAl2020}, we are not aware of a previous explicit treatment of evaporation-–accretion hysteresis\footnote{Hysteresis also occurs in physically distinct contexts of black hole physics, notably  hard-soft transition on accretion-state cycles of X-ray binaries \cite{MaccaroneCoppi2003,MeyerHofmeisterLiuMeyer2005}; and phase transition phenomena of anti-de Sitter black holes and their dual field theory observables \cite{CadoniFranzinTuveri2017,CadoniFranzinTuveri2017EPJC}.
The mechanisms in those cases are different from the retained-memory feedback studied here.} arising from an independently evolving retained-memory variable. Although accretion and memory-burdened evaporation have been considered together in the existing literature \cite{Maity2026,TabasiBerahmanFirouzjaee2023}, the resulting history-dependent multivalued luminosity at fixed mass is the particular feature we wish to illustrate here.

A minimal way to exhibit a genuine hysteresis is to let the black hole undergo a temporary accretion episode while the memory variable retains its own dynamics.
Using the same dimensionless variables defined for Eq.(\ref{eq:hillqdot}), we define the Hawking emission channel by
\begin{equation}
  \mathcal{J}_H(x,q)
  \equiv
  \frac{\Phi(q)}{3x^2}.
  \label{eq:grossHawkingchannel}
\end{equation}
A simple driven model that includes accretion is then
\begin{align}
\begin{cases}
  \dfrac{\d x}{\d \tau}
  &=a(\tau)-\mathcal{J}_H(x,q),
  \label{eq:hystx}\\
	\\
  \dfrac{\d q}{\d\tau}
  &=w\mathcal{J}_H(x,q)-\lambda q,
	\end{cases}
\end{align}
where $a(\tau)$ is a dimensionless accretion rate.
The crucial modeling choice is that Hawking emission writes the record through the \emph{gross} outgoing channel $\mathcal{J}_H$, whereas ordinary accretion changes the mass only without contributing to erasure of the memory. This is a natural choice: to undo the effect of a Hawking emission event would require the specially correlated microscopic time reverse of the emitted state, not merely generic incoming matter with the same energy.

Note that Eq.(\ref{eq:hystx}) is structurally similar to the Lotka-Volterra equation in mathematical ecology. However, due to a sign difference, in the latter system, more predators mean faster consumption of prey. In the memory-burden case, more memory deposited means \emph{slower} mass consumption\footnote{In addition, it can be shown that, within our considered model, constant positive accretion ($a>0$) and erasure ($\gamma > 0$) lead to a saddle equilibrium instead of the center type of the original Lotka-Volterra system.}. Thus our coupled system is still closer to product inhibition in chemical kinetics than to ordinary ecological prey-predator relationship\footnote{Another interesting comparison would be with the two-fluid dark energy toy models explored in \cite{2212.04429}, which is also of a generalized Lotka-Volterra-type.}.

\begin{figure*}[t]
\centering
\begin{subfigure}{0.48\textwidth}
    \includegraphics[width=\linewidth]{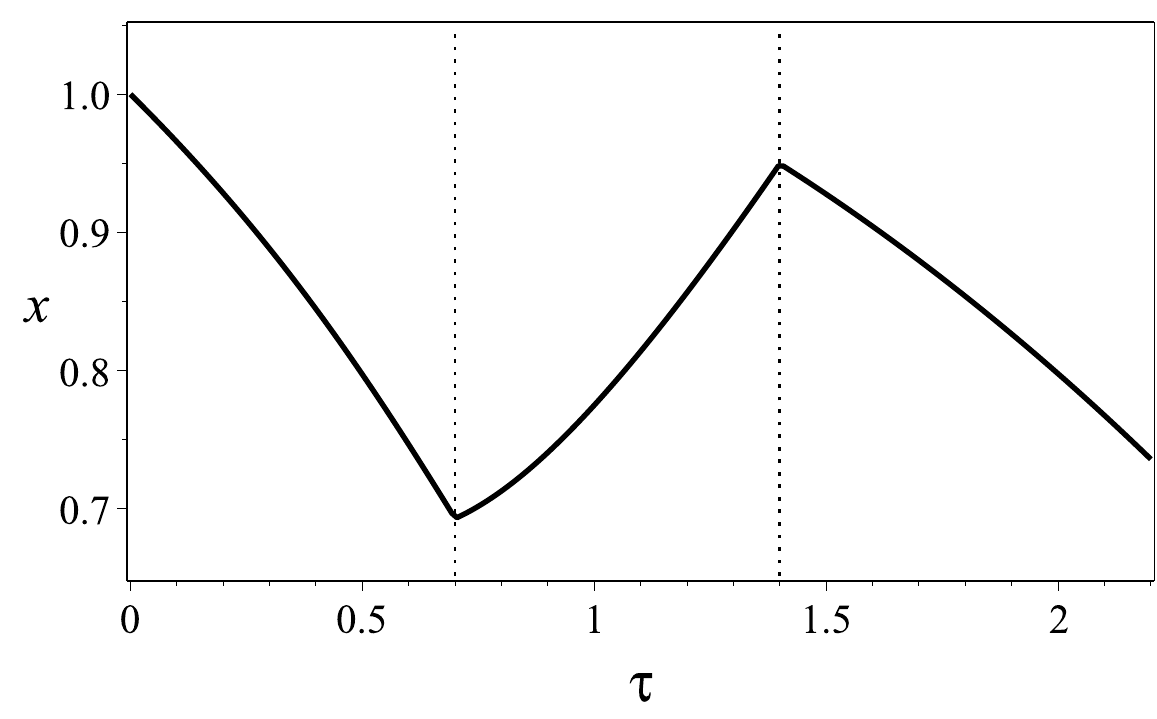}
\end{subfigure}
\hfill
\begin{subfigure}{0.48\textwidth}
    \includegraphics[width=\linewidth]{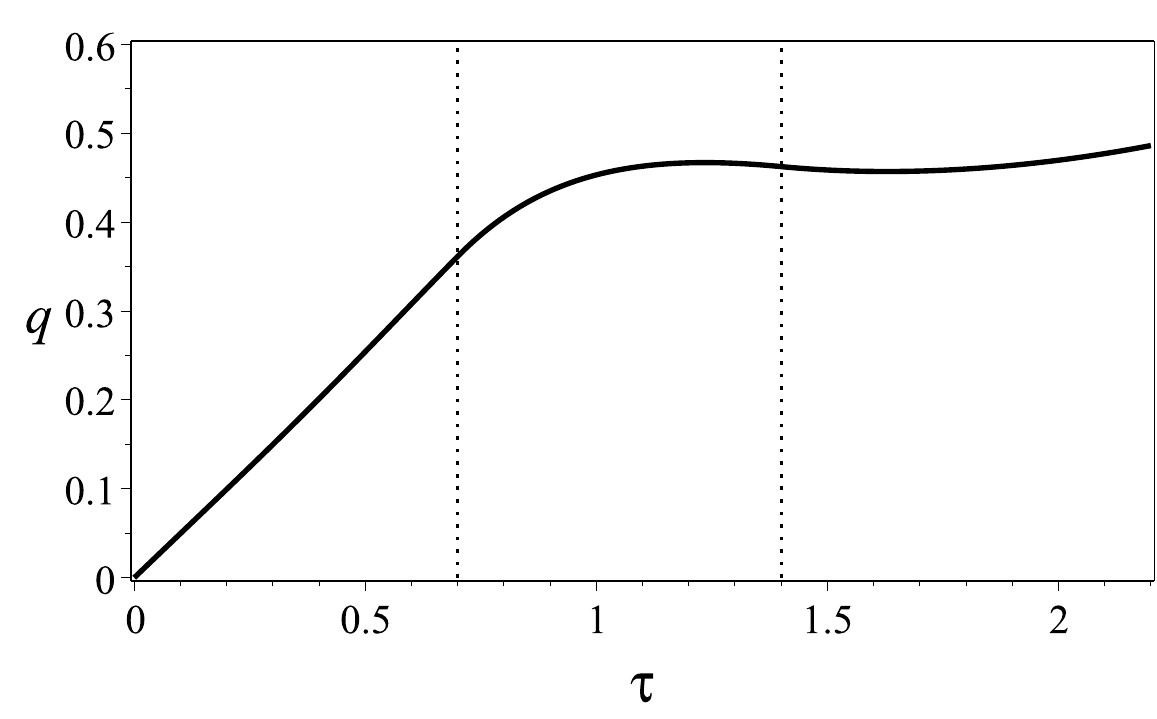}
\end{subfigure}

\medskip

\begin{subfigure}{0.48\textwidth}
    \includegraphics[width=\linewidth]{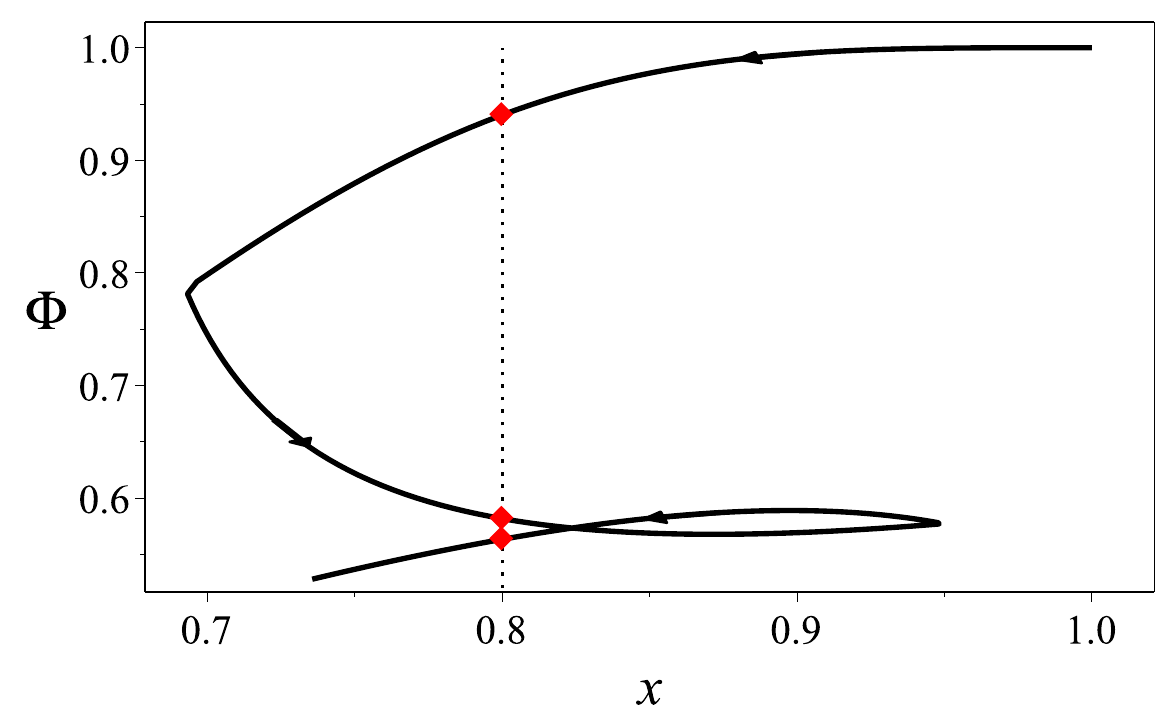}
\end{subfigure}
\hfill
\begin{subfigure}{0.48\textwidth}
    \includegraphics[width=\linewidth]{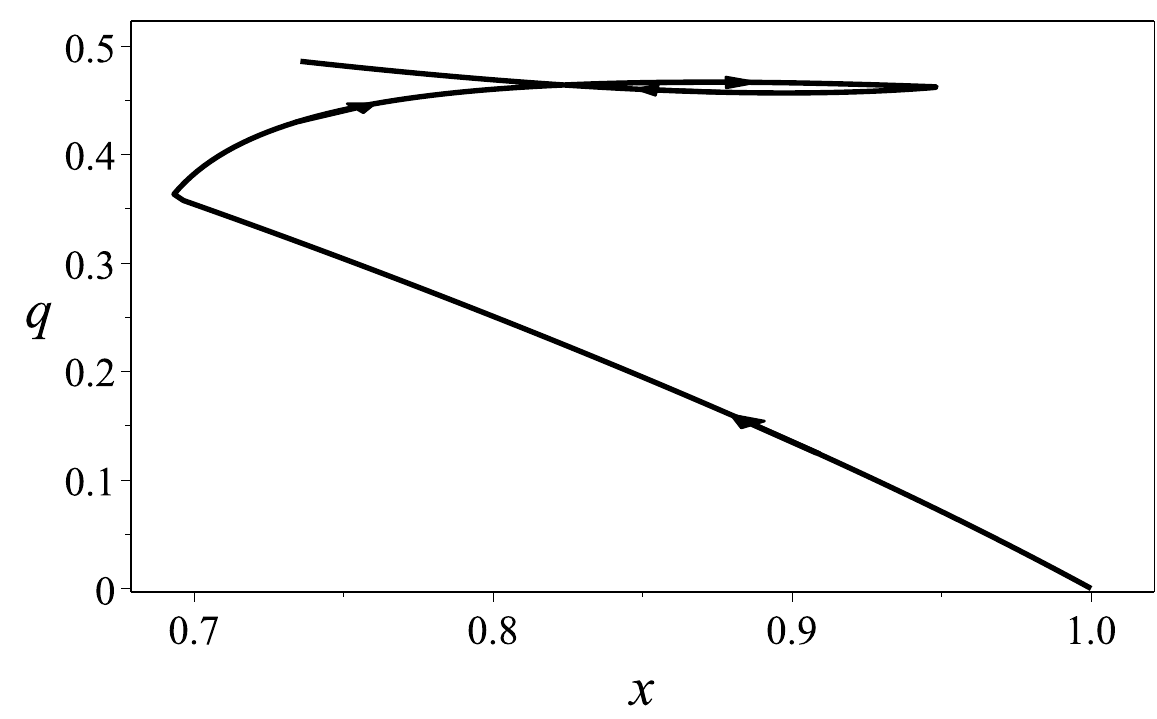}
\end{subfigure}

\caption{Toy hysteresis generated by Eqs.~\eqref{eq:hystx}--\eqref{eq:accretionpulse}.  \textbf{Top Left:} A temporary accretion episode reverses the net mass evolution. \textbf{Top Right:} The memory variable $q$ does not retrace the mass history because it is written by the gross Hawking channel and relaxes on its own timescale. For both the top figures, the vertical dotted lines mark the onset and termination of accretion at $\tau=0.70$ and $1.40$, respectively.  \textbf{Bottom Left:} Consequently the luminosity is multivalued as a function of mass: the same $M/M_0$ is encountered with different $\Phi = L/L_H$, the normalized luminosity. The red diamonds mark three passages through the same mass $x=0.8$, exhibiting distinct luminosity suppression factors. \textbf{Bottom Right:} In the full $(x,q)$ state space the trajectory forms a (non-closed) loop.  Arrows indicate the direction of evolution in both plots in the bottom figures. The model is only schematic.}
\label{fig:hysteresis}
\end{figure*}

For illustration, we take the Hill parameters to be
\begin{equation}
  n=4,
  \qquad q_c=0.5,
  \qquad w=1.5,
  \qquad \lambda=0.8,
\end{equation}
and a rectangular accretion pulse
\begin{equation}
  a(\tau)=
  \begin{cases}
    0.70, & 0.70\leqslant \tau<1.40,\\
    0, & \text{otherwise}.
  \end{cases}
  \label{eq:accretionpulse}
\end{equation}
The resulting trajectory is shown in Fig.(\ref{fig:hysteresis}).

It is observed that at the same mass $x\simeq0.8$, the first evaporation pass gives approximately
\begin{equation}
  q\simeq0.251,
  \qquad
  \frac{L}{L_H}\simeq0.940,
\end{equation}
whereas accretion increases the mass, resulting in
\begin{equation}
  q\simeq0.460,
  \qquad
  \frac{L}{L_H}\simeq0.582.
\end{equation}
On the subsequent evaporation pass, one finds $q\simeq0.469$ and $L/L_H\simeq0.563$ at the same mass. In other words, the luminosity is multivalued as a ``function'' of the mass. Within the kinetic-memory subclass two states can have the same mass and the same instantaneous Hawking temperature while radiating at different rates.

This is of course a very simple toy model. 
A more realistic implementation needs to first derive or motivate a specific form of the kinetic factor $\Phi$, but 
would also need to distinguish the outgoing Hawking spectrum, the types and correlations of accreted matter, or other environmental driving forces, as well as the specific microscopic erasure or rewriting channel rather than representing all of these effects simply by $a(\tau)$ and $\lambda$.
However the general expectation is that if the memory record has independent dynamics, a present-day mass need not determine a unique luminosity.
For an astrophysical population of black holes, this opens the possibility of branch structure or an intrinsic history-dependent spread in the emission properties at fixed inferred mass, rather than a universal single-valued luminosity-mass relation $L=L(M)$.

Let us remark that despite the loop that appears in the $\Phi$-$x$ plot in Fig.(\ref{fig:hysteresis}), it is not a closed loop (the technical term is that there is no ``return-point memory''). This is 
because $q$ evolves independently during the whole excursion, and thus writing and relaxation do not exactly undo one another when the mass returns.
In the no-erasure limit the non-closure is clearer to see explicitly since $\dot q =\eta L_w$ implies
\begin{equation}
q(t_2)-q(t_1) = \int_{t_1}^{t_2} \eta L_w \d t > 0,
\end{equation}
and consequently for a black hole that undergoes an evaporation-accretion cycle and returns to its original mass $M(t_2)=M(t_1)$, a strictly decreasing suppression function (since $q(t_2)>q(t_1)$) means that $L(t_2)<L(t_1)$ so the loop cannot close. 

A genuine closed hysteresis loop therefore requires an additional structure. For example, a Preisach-type construction \cite{Preisach1935,9210018} can be arranged to exhibit return-point memory, but only under the additional assumption that the accretion phase can somehow deactivate the kinetic influence of some previously activated memory sectors \footnote{In a Preisach-type description, an exact return-point closure requires that every effective sector switched during a nested excursion be restored to its previous state when the corresponding reversal point is revisited. If increasing the mass leaves all previously activated sectors kinetically active, the active set can still grow or at least remain unchanged, so returning to the same mass cannot in general restore the same $\Phi$.}. This does not mean that the accretion needs to \emph{undo} the memory writing process, but at the very least it needs to remove those memory sectors from the instantaneous feedback on the evaporation rate. Whether black-hole memory dynamics possesses such a structure is a separate microscopic question that we do not pursue here.

Another qualitatively different possibility arises in black-hole mergers. It has been argued that two memory-burdened black holes can merge into a ``young'' one that resumes essentially unsuppressed Hawking evaporation \cite{2410.07037,2506.13861,2607.03560}; see also \cite{2509.22540}. We propose that such a resetting of the effective memory state may be referred to as \emph{merger rejuvenation}, in parallel to existing terms in other astrophysical contexts; see e.g., \cite{0810.3902,1601.05084,2407.07937}. The blue straggler stars \cite{2410.10314} are especially interesting since one of the models for their formation involves a lower-mass star draining its larger-mass companion of hydrogen and thereby rejuvenating itself\footnote{Informally known as ``stellar vampirism''.}. In merger rejuvenation, it is the memory-writing capacity that undergoes a ``renewal''.

Unlike black hole evaporation considered thus far, a black hole merger amounts to a \emph{non-adiabatic} process. Therefore it is conceivable that such a process is accompanied by a rewriting, reorganizing, or resetting of the memory state, schematically,
\begin{equation}
(M_1,q_1)+(M_2,q_2)\longrightarrow(M_f,q_f),
\end{equation}
with $q_f$ no longer determined by adiabatic continuation of either progenitor. 

As argued in \cite{2511.01848}, for two burdened equal-mass black holes, the merged black hole has roughly twice the mass of one progenitor, hence four times its horizon area, providing substantially increased memory-storage capacity and thus no longer suppresses the Hawking emission as much as the progenitor. However, one must be cautious  about whether memory is stored on the horizon or via other means. In any case,
note that ``reset'' does not mean that information has been lost.
Quite the contrary, the post-merger state can preserve all the information and still possesses a large information load \cite{2607.03560}. The distinction between memory load and memory burden in the microscopic picture closely parallels (but is not equivalent to) the distinction between retention and feedback emphasized in our work: a large, persistent memory load need not modify evaporation unless the stored pattern becomes dynamically costly and couples back to the depletion process.

This ``memory-reset'' contrasts with the hysteretic evolution considered above, where the memory variable $q$ survives environmental driving and the system can revisit the same mass on a different branch.
Whether a merger erases, preserves, or reorganizes the accumulated memory is therefore a microscopic question about the retention and rewriting channels, and must be specified through additional phenomenological assumptions in the absence of a microscopic theory.

\section{Discussion}\label{Discussion}

We shall first look at how the results in this work may affect astrophysics of primordial black holes, after which, we will compare the memory-burden mechanisms in this work with other systems in many-body physics that display similar behaviors or mathematical structures. 

\subsection{Implications for PBH Phenomenology}
\label{sec:pbhphen}

Astrophysicists need not commit to a unique microscopic theory of black hole memory burden in order to phenomenologically investigate how memory burden may lead to observational signatures. 
One may simply specify a phenomenological evaporation law and state which structural assumptions it encodes.
For example, one may write the luminosity as
\begin{equation}
  L(M)=L_H(M) \Phi_\text{pheno}(M;M_0,x_{\rm on},\mathfrak{s},\ldots),
  \label{eq:phenPBH}
\end{equation}
where $\Phi_\text{pheno}$ is the phenomenological suppression factor, with $\Phi_\text{pheno}=1$ reproducing the standard Hawking rate.
The variable $x_{\rm on}$ denotes an onset mass fraction, while $\mathfrak{s}$ is a parameter that controls the strength of the suppression. More parameters may be included if so desired.
Eq.\eqref{eq:phenPBH} is perfectly adequate as a phenomenological ansatz, but by itself does not determine why the suppression occurs.
The present analysis suggests a minimal dictionary for interpreting such parameters.
The onset scale $x_{\rm on}$ asks when a sufficiently large history variable has been written; the persistence of the burden asks how that variable is retained; and the suppression factor asks how the stored history feeds back on the subsequent emission rate.

Our coupled ODE model Eq.(\ref{eq:phenq}) is of course not the first model proposed in the literature. Several prescriptions of this kind have been employed in the memory-burden literature. These include step-like transitions to entropy-suppressed evaporation ($\dot{M} \sim S^{-k} L_H$ for some constant $k$), and smooth hyperbolic-tangent interpolations between the early Hawking and burdened phases \cite{2503.21005}, as well as gradual transitions motivated by the growth of memory-mode energy gaps \cite{2503.21740}. The latter can produce an approximately exponential suppression over a finite fractional-mass interval, with a characteristic width related to the microscopic gap-growth exponent. Subsequent phenomenological studies have used the corresponding late-time suppression to investigate observational constraints on primordial black holes \cite{2506.13861,2512.06941}. 

These prescriptions establish useful and simple phenomenological models for the onset, width, and magnitude of evaporation suppression. Our Eq.(\ref{eq:phenq}), in addition, incorporates an independently evolving memory variable $q$ with an explicit writing and relaxation dynamics. This allows the effective suppression to depend on the evaporation history and, under external driving, to become multivalued at fixed mass. We remark that the existing different phenomenological prescriptions might correspond to different choices of $\Phi(q)$, the writing source $\mathcal{W}$, and the relaxation or erasure law $R[q]$. However, in view of the No-go Proposition, it is not guaranteed that every phenomenological prescription can be derived from a physically sensible memory dynamics.

An additional dynamical memory variable $q$ may appear unnecessary for many astrophysical applications, for which a phenomenological modification of the mass-loss rate is usually sufficient to calculate black hole lifetimes and the associated observational constraints. Indeed, when the memory variable is uniquely determined by the instantaneous mass, as in our erasure-free model with $q=w(1-x)$ in Eq.(\ref{eq:qexactHill}), the coupled system \emph{does} reduce to a single differential equation. We therefore do not advocate replacing the existing phenomenological prescriptions in astrophysical applications \emph{where such a description is adequate}. The advantage of the coupled formulation becomes apparent when memory possesses independent dynamics. In particular, accretion, relaxation, or other environmental processes can produce different retained-memory states at the same black hole mass, as illustrated by the hysteresis example in the preceding section. Such effects cannot be represented by a universal single-valued luminosity law depending only on the instantaneous mass, without introducing additional history-dependent parameters or prescriptions. The coupled approach thus provides a minimal dynamical extension of conventional phenomenological models, allowing the onset, retention, and feedback of memory burden to be specified and investigated separately. If an existing phenomenological model can be re-cast into the form of Eq.(\ref{eq:hillqdot}), then it admits an underlying write–retain–feedback interpretation and thus successfully evades the no-go result.

It should be noted that PBH observables are not determined by the lifetime alone.
The kinetic-memory subclass considered in our explicit toy model changes the overall emission rate while leaving the instantaneous Hawking temperature unchanged over the regime of validity.
A spectral-memory model may instead distort the relative occupation of emitted frequencies or particle species, while a genuinely thermodynamic-memory model may alter the effective temperature itself.
These possibilities can lead to different predictions for diffuse and point-source gamma rays, particle yields, plasma heating, Big Bang nucleosynthesis, dark radiation production, and other observables even when the total lifetime is comparable.
Recent phenomenological studies already illustrate this sensitivity: hot spot calculations, for example, explicitly compare different post-burden prescriptions and find qualitatively different thermal profiles \cite{LevyHeurtier2025}.
Likewise, gamma-ray searches and PBH survival studies necessarily inherit assumptions about the post-burden spectrum and mass-loss law \cite{Chianese2025,TanZhou2026,Maity2026}.

Hysteresis adds a further observable possibility.
If the state is genuinely described by $(M,q)$ rather than by $M$ alone, two objects with the same inferred mass need not have the same luminosity or spectrum.
A realistic search would therefore require more than a simple fitting of a universal curve $L(M)$. It would need to look for branch structure, delayed response to changing environments, or an intrinsic spread correlated with evolutionary history.

For practical use, a phenomenological PBH model can be characterized by at least four pieces of information: (1) the onset criterion for departure from Hawking evaporation; (2) the retention law of the history variable; (3) the feedback law controlling the mass-loss rate; and (4) the prescription for the emitted spectrum or effective temperature.
If hysteresis is allowed, one must also specify how accretion, mergers, or environmental driving write or erase the record. Of course, from the observational point of view, we do not know whether hysteresis is allowed or not, hence it might actually be more interesting to explore whether one can constrain the possibility of hysteresis from observational data if possible. This would, in return, provide valuable information for future theory-building. 

Of course, merger rejuvenation that we just discussed in the preceding section still remains an interesting arena to further explore the non-adiabatic properties of memory burden.

\subsection{Analogues From Many-Body Physics }

The memory-burden scenario was first motivated \cite{Dvali2018} by the quantum $N$-portrait picture, which explicitly describes a black hole as a critically packed Bose--Einstein condensate (BEC) of many weakly interacting soft gravitons and relates its nearly gapless Bogoliubov modes to information storage \cite{DvaliGomez2014}.
It is therefore instructive to ask what ordinary condensates or many-body systems, in addition to BEC, can teach us about the various stages of memory burden.

We note that bosonic Josephson junctions exhibit macroscopic quantum self-trapping, in which nonlinear interaction energy feeds back on tunneling strongly enough to inhibit large-amplitude population transfer. This effect has been observed directly in double-well condensates \cite{Albiez2005}.
Self-trapping is of course not the same as information burden. However, the dynamical structure is suggestive since it involves an internal collective state that changes the accessibility of subsequent transfer channels. Especially relevant to the present finite-versus-infinite lifetime distinction, exact finite-$N$ quantum dynamics need not preserve the mean-field trapped state forever (or as long as one's experiment permits). In fact a recent analysis finds breakdown of macroscopic self-trapping after a finite time for any finite particle number, with a long-lived quasi-self-trapped regime emerging at large $N$ in the symmetric two-mode Bose-Hubbard setting \cite{Bardin2026}.

A sharper parallel comes from the mean-field theory of structural glasses, where the infinite-dimensional limit provides an exactly solvable example \cite{ParisiUrbaniZamponi2020}.
In such mean-field descriptions the glass transition can produce ideal dynamical arrest and sharply defined metastable states, whereas in finite-dimensional systems activated hopping or nucleation can reopen escape channels.
Indeed, Biroli and Kurchan \cite{BiroliKurchan2001} emphasized that truly stable metastable states are generally artifacts of the mean-field approximation or the zero-temperature limit. In finite-dimensional systems, metastable states can decay through processes such as nucleation, so their physical meaning must be defined with reference to a finite timescale. A state may therefore appear effectively stable over an observational interval while remaining dynamically unstable on sufficiently long timescales. 

This is analogous to the distinction between a memory-burden model with an exact kinetic zero and one in which an additional channel leaves a tiny residual luminosity,
\begin{equation}
  L_{\rm tot}(M)=L_H(M)\Phi_{\rm mem}(M) +L_{\rm extra}(M),
  \label{eq:extraChannel}
\end{equation}
for which $\Phi_{\rm mem}(M_\star)=0$ does not imply an infinite lifetime if $L_{\rm extra}(M_\star)>0$.
Thus the mean-field/finite-dimensional distinction in glasses supplies a concrete precedent for separating exact arrest from effective stabilization. For astrophysical purposes, however, a black hole may be longevous without being eternal: its lifetime can exceed the age of the Universe by many orders of magnitude, rendering the distinction between finite and infinite lifetimes practically irrelevant for most observational applications.

There is also a striking mathematical parallel to the analysis of Sec.(\ref{MMO}) and Sec.(\ref{retention}).
The exact $d\to\infty$ dynamics of interacting particle systems can be reduced to an effective one-dimensional stochastic process with self-consistently determined retarded response and noise kernels \cite{MaimbourgKurchanZamponi2016,AgoritsasMaimbourgZamponi2019}.
In dynamical mean-field theory the kernels are determined self-consistently by the same effective stochastic process.
This makes the analogy with Eqs.\eqref{eq:kernel}--\eqref{eq:kernelvar} particularly useful (though of course the many-body kernels are determined self-consistently from underlying dynamics, whereas for our case $G(t,t')$ is only a schematic response kernel). 
Rapid decay of the retarded and correlation kernels corresponds to loss of dynamical memory and restoration of ordinary relaxation, whereas persistent long-time kernel support is associated with dynamical arrest.
The microscopic variables, of course, are entirely different from those of an evaporating black hole, but the mathematical lesson is transferable: long-lived macroscopic memory is encoded not merely by the existence of many microscopic degrees of freedom, but by the temporal support of the response and correlation structures that feed their history back into later dynamics.

Kinetically constrained models offer another useful comparison to our kinetic-freezing scenario. These systems can have non-interacting equilibrium thermodynamics while displaying slow relaxation, aging, and dynamical arrest because the allowed transitions are constrained \cite{RitortSollich2003}.
Prethermal many-body systems under a sufficiently high-frequency periodically driven setting provide a further caution, since local dynamics can suppress energy absorption exponentially and generate parametrically long-lived metastable states without producing exact stability \cite{AbaninEtAl2015}. 

Again, we should emphasize that these parallels do not constitute models of black-hole memory, but they do reinforce two structural lessons: (1) stabilization is a statement about transition dynamics, and (2) an enormous lifetime is not the same as a truly infinite one.

\section{Conclusion: When Does Hawking Evaporation Remember Its Past?}\label{Conclusion}

In this work, we have examined under what conditions an evaporating black hole can be memory-burdened. We started with a simple stochastic analysis of Hawking evaporation that led to a no-go result, from which we identified what conditions may be broken to achieve memory burden. This was followed by some concrete, albeit toy, examples that illustrate some properties of memory-burdened black holes, including the possibility that accretion together with Hawking emission can give rise to a hysteresis phenomenon such that black holes with identical masses may have different luminosities.  

We shall conclude this work by noting that
the logical distinction between information storage and dynamical feedback is reminiscent of Maxwell's demon. The demon may measure a system and retain a perfectly good record while never using that record to operate the door that separates the hot and cold molecules. Likewise, a black hole may preserve microscopic information without requiring that information to enter the subsequent evaporation.
In the language of this paper, the demon must not merely write and retain, it must also ``feed back'' and act on that information.
A memory-burden model makes the door mechanism sensitive to the stored record, whereas a no-burden construction can keep the ``notebook'' without allowing it to influence the system at all.

In addition, Landauer's principle provides the familiar example that logically irreversible erasure of one bit in an ordinary thermal memory has a minimum heat cost $Q_{\rm erase}\geqslant k_B T\ln2$ in the ideal limit \cite{Landauer1961}. We neither identify black hole memory relaxation with Landauer erasure, nor assume that a coarse-grained variable $q$ literally counts classical bits.
Nevertheless, the act of writing, retaining, erasing, reorganizing and using a record to realize feedback are all distinct physical operations.
Accordingly, a phenomenological erasure term such as $\mathcal{R}[q]=\gamma q$ should ultimately be replaced by an actual transfer, scrambling, or loss of dynamical accessibility in some concrete manner in a specific microscopic or more fundamental theory. Some physical laws akin to Landauer's principle may play an essential role in all this.

Nevertheless, for astrophysical purposes, one often works with simple phenomenological models. In this work we gave two toy examples in the kinetic-memory subclass, namely the Hill model that prolongs the black hole lifetime (but still ends with a complete evaporation) and the terminal feedback model that terminates at a finite mass (but only asymptotically in time). For more general and perhaps more well-motivated models, our work suggests some important steps to consider in the model-building:
\begin{itemize}
\item[(1)] Identify what writes the history;
\item[(2)] Determine whether the writing is systematic or fluctuating;
\item[(3)] Quantify how long correlations and retention persist;
\item[(4)] Specify how effective erasure or environmental driving acts;
\item[(5)] Show that the retained variable feeds back strongly enough to change the evaporation.
\end{itemize}

To answer the question posed in the subtitle of this section: Hawking evaporation remembers its past only when the three requirements---\textsc{write}, \textsc{retain}, and \textsc{feedback}---are all satisfied on the relevant dynamical scales. The last step is necessary in order for memory burden to be realized, otherwise information can be retained without necessarily affecting the black hole evaporation rate.

The black hole admits dynamical memory only when the demon actually opens the door.
\begin{quote}
\emph{Oblitus manet, nisi daemon ianuam aperiat.}
\end{quote}

\begin{acknowledgments}
This research is supported by NUAA funding No.1018-ILF26028.
\end{acknowledgments}

\end{document}